\documentclass[preprint]{revtex4-2}

\usepackage[utf8]{inputenc}
\usepackage{graphicx,subfig}
\usepackage{siunitx}
\usepackage{xcolor}
\usepackage{colortbl}
\usepackage{multirow}
\usepackage{amsmath}
\usepackage[bb=dsserif]{mathalpha}
\usepackage{bm}
\usepackage{bm}
\usepackage[normalem]{ulem}
\usepackage{xfrac}
\usepackage{soul}
\usepackage{todonotes}

\def\I{\text{i}}
\def\d{\text{d}}
\def\euler{\text{e}}
\def\eA{|\,\mathbf{e},\sfrac{1}{2}\rangle}
\def\eB{|\,\mathbf{e},\sfrac{3}{2}\rangle}
\def\eC{|\,\mathbf{e},\sfrac{5}{2}\rangle}
\def\gA{|\,\mathbf{g},\sfrac{1}{2}\rangle}
\def\gB{|\,\mathbf{g},\sfrac{3}{2}\rangle}
\def\gC{|\,\mathbf{g},\sfrac{5}{2}\rangle}

\begin{document}

\title{Photon echoes using atomic frequency combs in Pr:YSO --- experiment and semiclassical theory}

\author{Aditya~N.~Sharma$^1$}        
\author{Zachary~H.~Levine$^{2}$}   
\author{Martin~A.~Ritter$^1$}        
\author{Kumel~H.~Kagalwala$^1$}                       
\author{Eli~J.~Weissler$^{2,3}$} 
\author{Elizabeth~A.~Goldschmidt$^{4,5}$} 
\author{Alan~L.~Migdall$^{1,2}$}  
\affiliation{
$^\text{1}$Joint Quantum Institute, University of Maryland, College Park, MD 20742, USA
}
\affiliation{
$^\text{2}$Quantum Measurement Division, National Institute of Standards and Technology, Gaithersburg, MD 20899, USA
}
\affiliation{
$^\text{3}$Department of Electrical, Computer, and Energy Engineering, University of Colorado, Boulder, CO 80309, USA
}
\affiliation{
$^\text{4}$Department of Physics, University of Illinois, Urbana, IL 61801, USA
}
\affiliation{
$^\text{5}$U. S. Army Research Laboratory, Adelphi, MD 20783, USA
}
\date{\today}

\begin{abstract}
Photon echoes in rare-earth-doped crystals are studied to understand the challenges of making broadband quantum memories using the atomic frequency comb (AFC) protocol in systems with hyperfine structure.
The hyperfine structure of Pr$^{3+}$ poses an obstacle to this goal because frequencies associated with the hyperfine transitions change the simple picture of modulation at an externally imposed frequency.
The current work focuses on the intermediate case  where the hyperfine spacing is comparable to the comb spacing, a challenging regime that has recently been considered. Operating in this regime may facilitate storing quantum information over a larger spectral range in such systems.

In this work, we prepare broadband AFCs using optical combs with tooth spacings ranging from 1~MHz to 16~MHz in fine steps, and measure transmission spectra and photon echoes for each. We predict the spectra and echoes theoretically using the optical combs as input to either a rate equation code or a density matrix code, which calculates the redistribution of populations. We then use the redistributed populations as input to a semiclassical theory using the frequency-dependent dielectric function.
The two sets of predictions each give a good, but different account of the photon echoes.

\end{abstract}

\maketitle










\section{Introduction}
\label{sec:intro}

Quantum memory is a crucially important technology. Perhaps most prominently, it will play an important role in long-range quantum networks~\cite{Acin2018,Pompili2021}, which will be required to fully realize the power of quantum computers and other quantum technologies: quantum repeaters will be necessary to overcome transmission losses in these networks~\cite{Briegel1998}, and most current approaches to quantum repeaters depend on optical quantum memory~\cite{Duan2001,Awschalom2018},
a device for storing quantum states and recovering them at a later time. A practical quantum memory would enable long-lived storage and high-fidelity, high-efficiency retrieval of broadband single photons.
Rare-earth-ion-doped crystals are promising candidates for implementing quantum memory due to their long ground-state lifetimes, large inhomogeneous bandwidths, and narrow homogeneous bandwidths~\cite{simon2007quantum}. Various approaches to quantum memory have been developed for rare-earth-crystal platforms. The atomic frequency comb (AFC) protocol is one such scheme~\cite{Afzelius2009}. The central idea is to store single photons in a material with a frequency-comb absorption spectrum, with comb-tooth spacing $f_{\rm rep}$, and then to recover photon echoes after a time $T_{\rm rep}=1/f_{\rm rep}$.

Pr$^{3+}$:Y$_2$SiO$_5$ (Pr:YSO) is suitable for various optical storage techniques, and it was the first material used to demonstrate on-demand AFC quantum memory~\cite{Afzelius2010}. It has been studied extensively, including work on photon
echoes~\cite{Graf1997,Graf1998}, spectral-hole burning~\cite{Nilsson2004}, optical filtering~\cite{beavan2013demonstration}, electromagnetically induced transparency~\cite{Fan2019}, and stimulated Raman adiabatic passage~\cite{Klein2007,Gao2007}. The latter technique was used for selective retrieval of pulses~\cite{Wang2008}, and further studies investigated the rephasing of these coherent populations~\cite{Mieth2012} and on-demand retrieval based on spontaneous Raman scattering~\cite{Goldschmidt2013}. Storage of optical pulses has been demonstrated with stopped light using electromagnetically induced transparency~\cite{Longdell2005,Heinze2013} and a memory has been realized at the single-photon level using stopped light in a spectral hole~\cite{Kutluer2016}. Gradient-echo quantum memory was demonstrated~\cite{Hedges2010}, and AFC quantum memory has also been demonstrated at the single-photon level~\cite{Rielander2014,Seri2019}.
Recently, Pr:YSO has been used for high-rate entanglement distribution as a step towards the implementation of quantum repeaters \cite{Mannami2021} and on-demand quantum memory~\cite{Horvath2021}.

The AFC protocol has also been studied in other rare-earth-ion-doped crystals, which have similarly advantageous optical properties. Ref.~\cite{Bonarota2010} studied the effect of comb-tooth shape on echo efficiency. Storage and retrieval of single photons has been demonstrated: time-bin qubits~\cite{Usmani2010, Davidson2020} and entanglement with another photon~\cite{Saglamyurek2011} storage.
Temporal multimode storage has been studied~\cite{Jobez2016}. Pulse storage and retrieval has also been demonstrated using Stark shifts~\cite{Zhong2017}, superhyperfine levels~\cite{Askarani2019} and hybridized electron-nuclear hyperfine levels~\cite{Businger2020}. Gigahertz bandwidths were demonstrated using Tm$^{3+}$~\cite{Saglamyurek2011} and Er$^{3+}$~\cite{Saglamyurek2016} ions. Er$^{3+}$ has also been studied for use in quantum repeaters~\cite{Craiciu2021}.

One attractive feature of the AFC protocol highlighted in Ref.~\cite{Afzelius2009} is that in principle it enables retrieval efficiency arbitrarily close to unity. In practice, however, experimental demonstrations have achieved limited efficiency: Refs.~\cite{Sabooni2013,Jobez2014} used cavity enhancement to reach efficiencies just over 50\%.

Many rare-earth species and isotopes, including Pr:YSO, have hyperfine structure, which presents an obstacle to preparing high-bandwidth AFCs. To circumvent this challenge, past works on AFCs in Pr:YSO have followed two approaches: (1) use AFC bandwidth smaller than the hyperfine spacing \cite{Goldschmidt2013,Rielander2014,Kutluer2016}, or (2) use AFC tooth spacing larger than the entire 37~MHz spectral width of the hyperfine levels~\cite{Nicolle2021}. (Ref.~\cite{Seri2019} used a combination of these methods.) Here, we explore the intermediate regime in which the AFC bandwidth is larger than 37~MHz and the AFC tooth spacings are comparable to the hyperfine spacing. 
While some past studies on AFCs worked in this regime~\cite{Askarani2019,teja2019photonic}, here we systematically investigate AFC formation and echo retrieval for a range of comb-tooth spacings. We experimentally demonstrate AFC formation and echo retrieval across a range of comb-tooth spacings.
Through this systematic study, we hope to elucidate the dynamics between AFC bandwidth and hyperfine structure, which could aid in the design of future high efficiency AFC-based quantum memories.

We also develop a semiclassical theory of photon echoes which calculates the effect of the experimental optical fields applied to the crystal.  Historically, photon echo phenomena have been understood as collective emission by atoms prepared in a Dicke state~\cite{Dicke1954}, whereas here we achieve good results with a semiclassical model in which individual Pr$^{3+}$ ions, treated quantum mechanically, interact with a classical electromagnetic field.
Although several authors have noted the presence of a semiclassical limit~\cite{Sangouard2007,Gorshkov2007b,Afzelius2009}, and the formalism of this limit was presented in a recent review article~\cite{Chaneliere2018}, there does not appear to be a realistic calculation of photon echoes for rare-earth-doped crystals in the literature, particularly not one which calculates the AFC based on an observed optical signal as well as the photon echoes.
We calculate the atomic population shifts that occur during AFC preparation; the change to the dielectric response due to these shifts; and the echoes that arise from a pulse interacting with the highly dispersive medium.  We present two sets of results, one corresponding to AFCs as predicted by the rate equations and a second set as predicted by the density matrix formalism.  Given the redistribution of population which is a quantitative description of the AFC, the photon echo is described within the same formalism of highly dispersive linear optical response.

Regarding other calculations in this area, Theil et al.~\cite{thiel2014measuring} gave a theoretical account matching experimental results of pulse echo linewidths in three rare-earth crystals Tm$^{3+}$:YAG, Tm$^{3+}$:LiNbO3 and Tm3$^{3+}$:YGG.  More frequently, the calculations are models~\cite{xiong2008numerical} or done to propose experiments~\cite{arslanov2017optimal,tittel2010photon} rather than to do a careful comparison.



\begin{figure}
\centering
\includegraphics[width=6.35in]{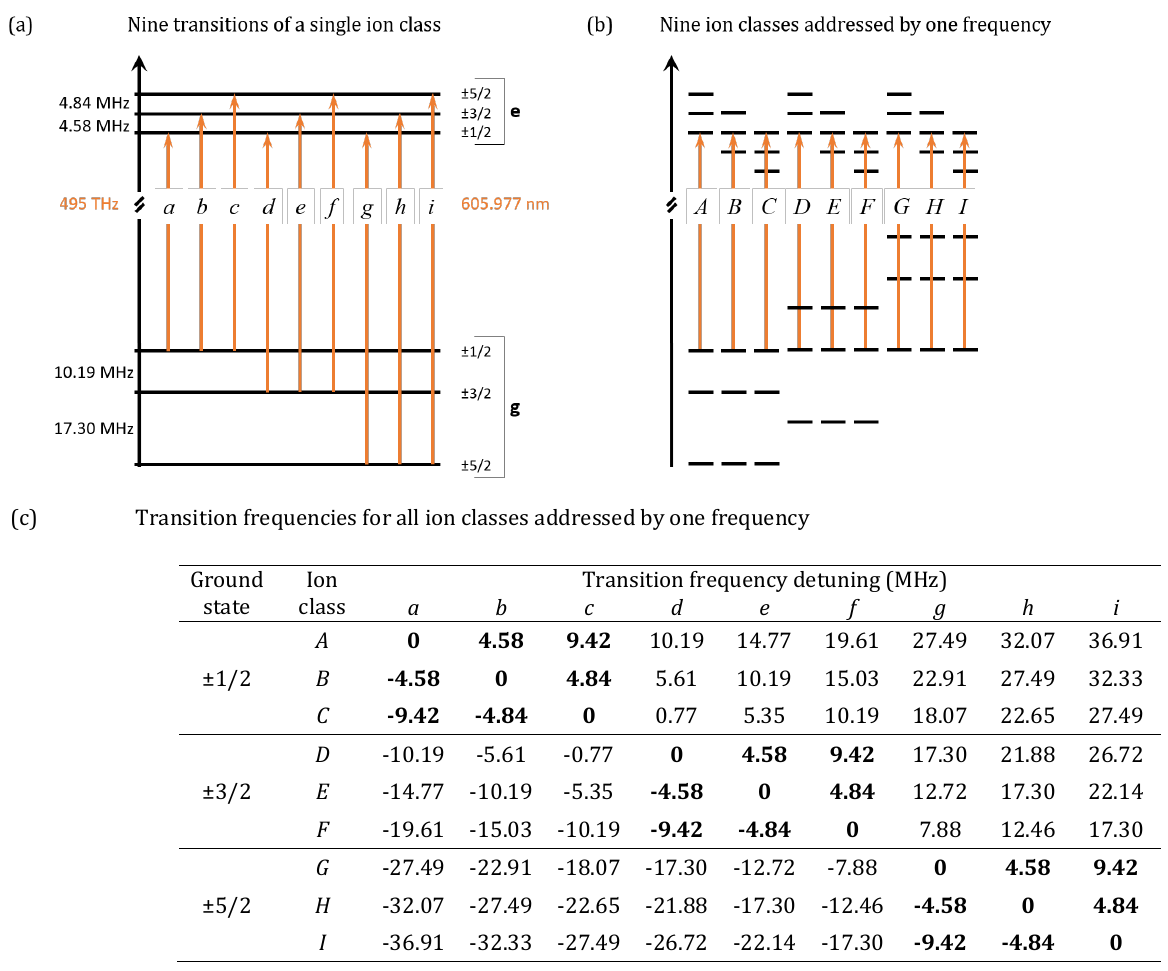}
\caption{Each Pr$^{3+}$ ion in Pr:YSO has 6 excited states and 6 ground states, all of which are two-fold degenerate in the absence of an external magnetic field. (a) Thus, each ion has 9 different transition energies $a$-$i$. (b) Conversely, any laser frequency within the 4.4~GHz inhomogeneous bandwidth addresses one of these transitions for 9 different ion classes $A$-$I$. (c) Table of transition frequencies for all ion classes addressed by a single laser frequency at zero detuning. Each row shows all 9 transitions for one ion class. Rows are grouped by common ground states. Since the laser frequency addresses transition $x$ for class $X$, the diagonal elements are zero. The bold-face diagonal-block elements indicate transitions with depopulated ground states, leading to peaks in the transmission spectrum in Fig.~\ref{fig:singleFreqResponse}, while the other elements indicate troughs. The bold-face elements depend only on the excited-state spacings, while the others also depend on the ground-state spacings.}
  \label{fig:introFigure}
\end{figure}

\begin{figure}
\centering
\includegraphics[width=6.35in]{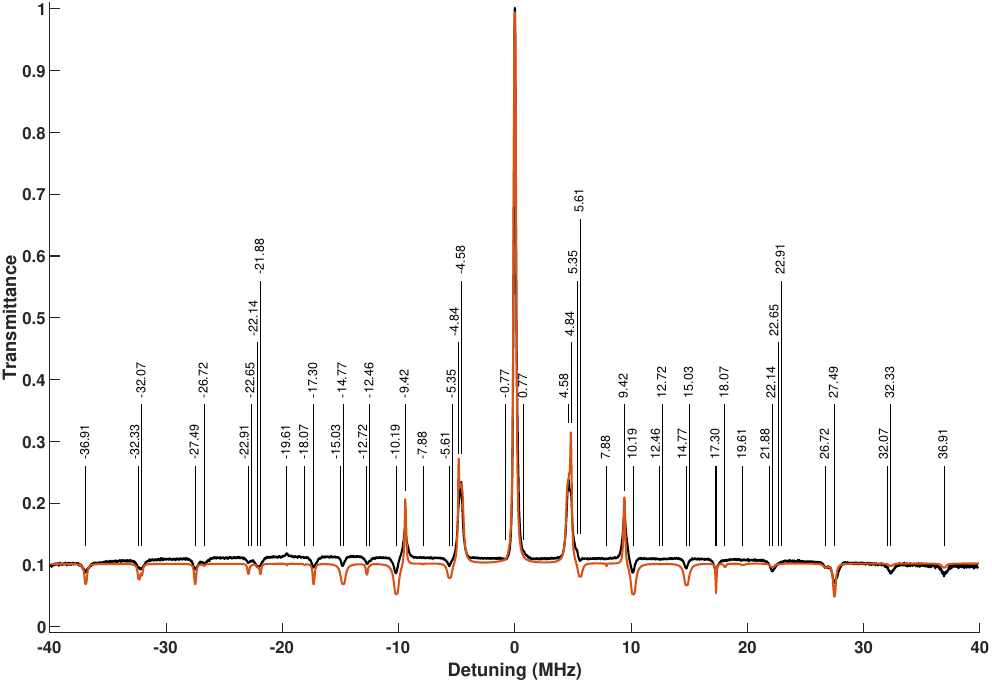}
\caption{Pr:YSO transmission spectrum after spectral-hole burning at zero detuning, as measured experimentally (black) and as predicted by density matrix calculation (red). The curves have been normalized to unit transmittance. The transition frequencies from Fig.~\ref{fig:introFigure}c are shown here.
}
  \label{fig:singleFreqResponse}
\end{figure}

The YSO crystal structure has two distinct Y sites where Pr$^{3+}$ impurities can occur. Here we study Pr$^{3+}$ ions at site~1~\cite{Maksimov1969crystal,Holliday1993}, which have $^3H_4\,\rightarrow\,^1D_2~(\mathbf{g}\rightarrow\mathbf{e})$ optical transitions at 605.977~nm. The homogeneous linewidth of these optical transitions is less than 1~kHz and can be observed when the crystal is cooled to liquid helium temperatures to eliminate phonon broadening; the inhomogeneous broadening, arising from local variations in the crystal structure surrounding each Pr$^{3+}$ ion, is typically several gigahertz~\cite{Equall1995}.
In the absence of an external magnetic field, the hyperfine states are all two-fold degenerate, and the ground- and excited-state manifolds each have three distinct energy levels (Fig.~\ref{fig:introFigure}). Therefore, each ion has nine transition energies labeled $a-i$ in ascending order; conversely, for any laser frequency within the inhomogeneous bandwidth, there are nine different classes of ions $A-I$ for which one of these transitions is resonant with the laser, with the laser addressing transition $x$ for class $X$.

As a result of the hyperfine structure, spectral-hole burning at even a single frequency results in multiple features
(Fig.~\ref{fig:singleFreqResponse}). The large transmittance peak at zero detuning occurs because all 9 classes become transparent at this frequency. The smaller peaks occur because some subset of the classes become transparent: for example, only 3 of the 9 classes that would normally absorb at 9.42~MHz detuning are depopulated. Due to power broadening, our measurement cannot distinguish features separated by less than 0.2~MHz, but such closely-spaced peaks may appear as broader peaks or shoulders. The transmission peaks in Fig.~\ref{fig:singleFreqResponse} correspond to population depletion at detuning frequencies shown in bold in Fig.~\ref{fig:introFigure}c.  The other frequencies correspond to population enhancement, hence transmission dips in most cases.

The excited-state spacings in Pr:YSO are nearly equal, leading to a comb-like response to spectral-hole burning at a single frequency. This occurs because (1) the transparency for classes $A,D,G$ at 4.58~MHz detuning overlaps with the transparency for $B,E,H$ at 4.84~MHz, and (2) an additional transparency occurs for $A,D,G$ at 9.42~MHz, further extending the comb-like response (similar reasoning for classes $C,F,I$ and $B,E,H$ explains the peaks at $-9.42$~MHz, $-4.84$~MHz, and $-4.58$~MHz).
This observation suggests that the material has a naturally periodic spectrum. Although it has been noted in previous works, for example Ref.~\cite{Askarani2019}, that this periodicity dictates which tooth spacings are compatible with AFC formation, here we carefully examine tooth spacings close to the periodicity. In fact, we find that AFC formation is optimal \textit{not} for tooth spacings exactly matching the excited-state hyperfine spacings, but rather a few hundred kilohertz detuned from those values.

\section{Experiment}
\label{sec:expt}
We use our experimental setup for two different measurement protocols, a comb-measurement protocol and a pulse-echo protocol, each of which proceeds in two stages. In the first stage of both protocols, we prepare an AFC. In the second stage, we either probe the transmission spectrum (the comb-measurement protocol), or observe the transmission and echoes of a pulse sent to the AFC (the pulse-echo protocol).

\begin{figure}
    \centering
    \includegraphics[width=15cm]{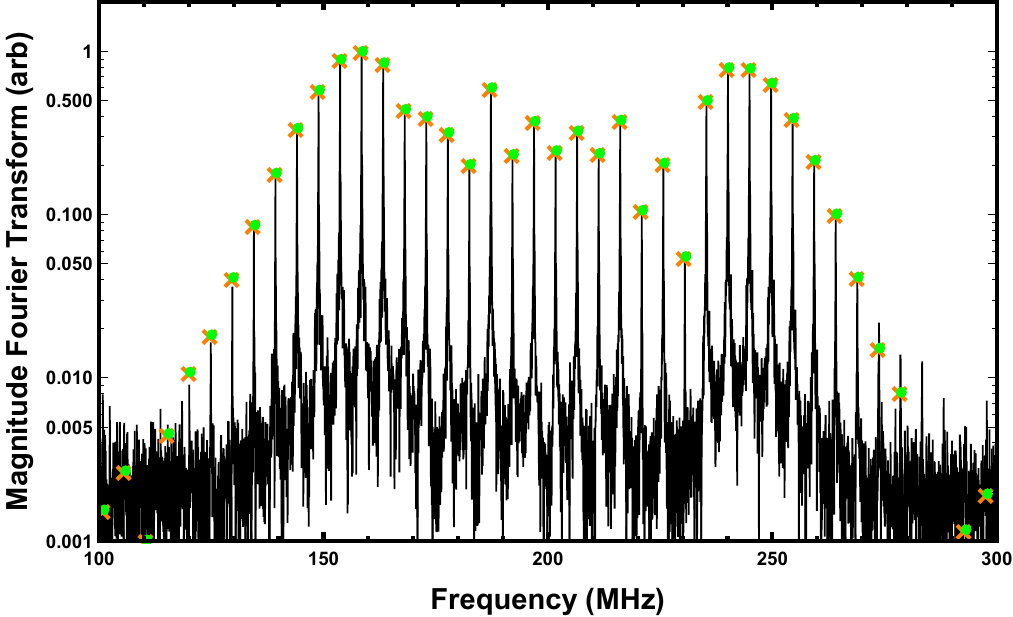}
    \caption{Optical frequency comb generated by AOM frequency modulation and observed by heterodyne mixing with a local oscillator (black curve). Here, the modulation frequency, and thus the tooth spacing, is 4.8~MHz. Our theoretical calculations require averaging, as described in the Appendix, to obtain a strictly periodic approximation to this signal.  Also shown is the Fourier transform of the signal (orange points) of Fig.~5  and the transform with a small shift required to achieve phase periodicity (green points).}
    \label{fig:fourierComparison}
\end{figure}

We use a tunable diode laser near 1212~nm, which is amplified in a fiber Raman system and then doubled, producing a beam at the $^3H_4\rightarrow{^1}\!D_2$ optical transition. To stabilize the laser frequency, we use Pound-Drever-Hall feedback~\cite{Black2001} to lock the infrared light  to a reference cavity with a Zerodur (Schott Glass),  spacer, held in a vacuum chamber. This beam is modulated using double-passed acousto-optic modulators (AOMs) to prepare the three different beams used in our measurements. A burn beam is produced by applying a frequency-modulated (FM) radio-frequency (RF) sine wave to the . This signal has a frequency-comb structure, centered at 100~MHz, with 60~MHz bandwidth, and a tooth spacing equal to the FM modulation frequency. After the double pass, this results in an optical frequency comb with 120~MHz bandwidth and tooth spacing equal to the FM modulation frequency (Fig.~\ref{fig:fourierComparison}). A probe beam is produced by (slowly) sweeping the AOM frequency, resulting in a frequency-swept beam covering the same bandwidth as the burn beam. Alternatively, 10~ns pulses are generated by applying RF pulses to the AOM.

\begin{figure}
    \includegraphics[width=5in]{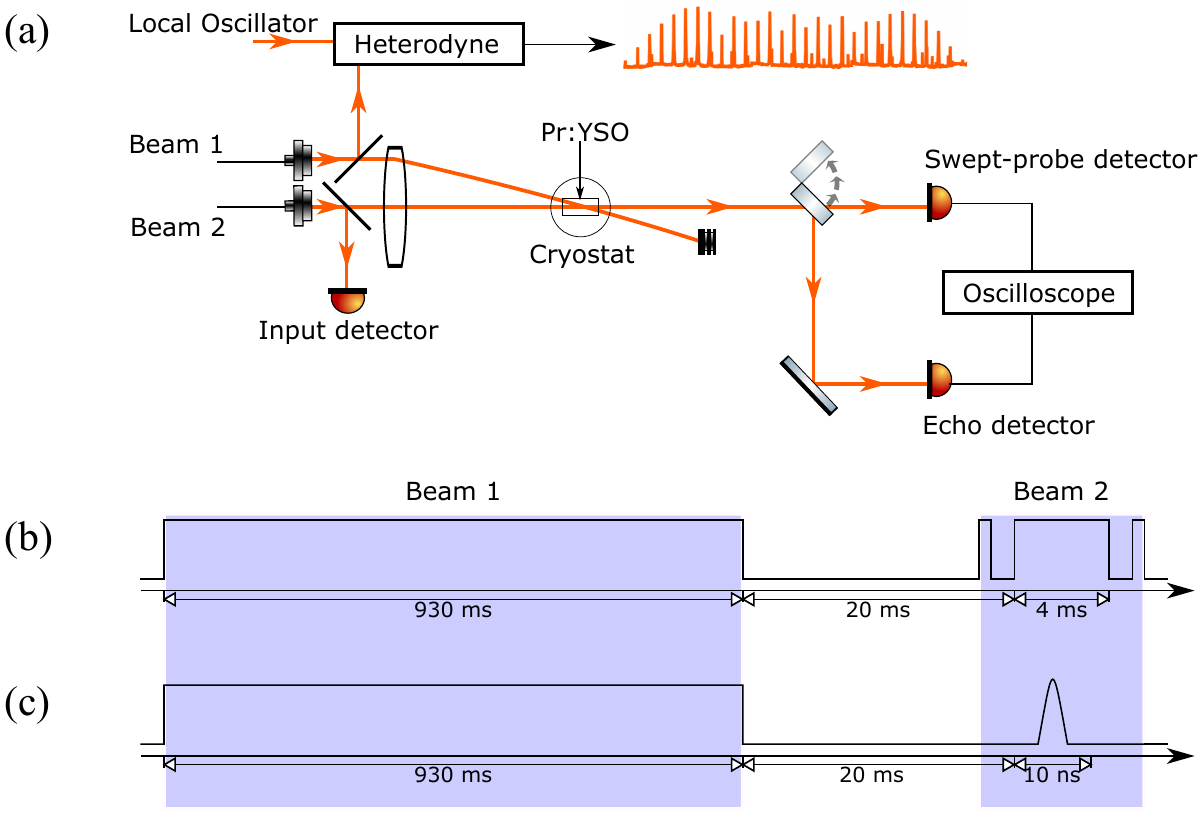}
    \caption{(a)~We prepare an AFC by burning the crystal for 30~seconds with an optical frequency comb (Beam 1). We pick off some of this beam to beat against a local oscillator for a balanced heterodyne measurement of the optical spectrum with which we drive the crystal. We use Beam~2 to make two different measurements (see text). Both measurements are repeated for approximately 30~cycles and the results are averaged. Each cycle is synchronized with the cryostat pump cycle to minimize the effect of vibrations. (b)~Timing diagram for one cycle of the AFC measurement (not to scale). During the 4~ms probe period, the laser frequency is swept across the AFC bandwidth to measure transmission. The short pulses on either side of the probe are used to demarcate the sweep duration to help automate data analysis; to minimize their effect on the measurements, they are tuned to a frequency 20~MHz outside the AFC and sweep bandwidths. (c)~Timing diagram for one cycle of the echo measurement (not to scale).
    }
    \label{fig:expLayout}
\end{figure}

In our setup (Fig.~\ref{fig:expLayout}), we use a Pr:YSO crystal with 0.05\% substitution of Pr for Y, housed in a cryostat maintained below 4~K. Two beams are focused to overlapping waists inside the crystal. Beam~1 is the burn beam described above: we vary the tooth spacing of this optical frequency comb in 0.1~MHz steps, from 1.0~MHz to 16~MHz. The first stage of both measurement protocols consists of 30~seconds of nearly continuous burning with Beam~1, aiming to imprint the optical frequency comb on the atomic transmission spectrum and establish an AFC. In our comb-measurement protocol, Beam~2 is the probe beam described above.  During the 4~ms frequency sweep of the probe beam, we acquire the transmission spectrum of the crystal. In the pulse-echo protocol, Beam~2 is the 10~ns pulse described above. We monitor emission from the crystal to detect the transmitted pulse and its echoes. Due to the different natures of the output signals from the two measurements, detectors with different response bandwidths are required. In the second stage of each measurement protocol, the 1~s measurement cycle is synchronized to an electrical pulse indicating the cryostat pump cycle, in  order to minimize effects of vibrations. We repeat the measurements for approximately 30 cycles~\cite{note:cycleNumber} and average the results. In addition to the 30~s burn in stage~1, in stage~2 of each cycle, 930~ms is spent burning the crystal to reinforce the AFC.

\section{Theory}
\label{sec:theory}

Most discussions of the AFC protocol rely on preparation of a Dicke state of the atomic ensemble~\cite{Sangouard2007,Gorshkov2007b,Afzelius2009,Afzelius2010,Minar2010,Jin2015,Jobez2016,Saglamyurek2016,Askarani2019,teja2019photonic,Furuya2020,Holzapfel2020,Etesse2021,Nicolle2021}; in this paper, however, we show that the AFC transmission spectra and echo pulses can be obtained using only semiclassical concepts, provided the beams are in coherent states.
Not all authors invoke the Dicke state.
Chaneli{\`e}re et al.~\cite{chaneliere2010efficient} also outlines a semiclassical treatment of photon echoes in AFCs, but the calculation of the susceptibility is not explicit.
Burman and Le~Gouet~\cite{berman2021pulsed} have made an analysis of both uniform and random AFCs in highly dispersive media using the Maxwell-Bloch equations for a two-level system, with an emphasis on analytic modelling.

In Section~\ref{subsec:refIndex}, we describe the theory required for calculating the dielectric response function. In Section~\ref{subsec:propagation} we show that this function can be used to calculate the propagation of light pulses through the crystal. Finally, in Section~\ref{subsec:denMat} we describe our method for determining how an incident field changes the ground-state population distribution.

\subsection{Calculating the dielectric response function}
\label{subsec:refIndex}

The Clausius-Mossotti relation gives the macroscopic dielectric function $\varepsilon(\omega)$ of a solid in terms of the polarizabilities $\alpha_k(\omega)$ of its constituent atoms. The relation remains valid for particles embedded in a homogeneous medium if $\varepsilon(\omega)$ is interpreted as the dielectric constant relative to the undoped crystal dielectric constant $\epsilon_0\varepsilon_{\rm bkg}$~\cite{Levy1992}:
\begin{eqnarray}
3\frac{\varepsilon(\omega)-1}{\varepsilon(\omega)+2}
= \frac{1}{\epsilon_0 \varepsilon_{\rm bkg}}
\sum_k N_k \alpha_k(\omega)
.\label{eq:Clausius}
\end{eqnarray}
Here, the index $k$ denotes different species of oscillators, $N_k$ is the number density of such oscillators, $\epsilon_0$ is the vacuum permittivity, and $\omega$ is the frequency~\cite{Mohr2015}. When applied to optical frequencies, Eq.~(\ref{eq:Clausius}) is often called the Lorentz-Lorenz relation~\cite{zangwill2013modern}. In our problem, values of $k$ represent the ion classes forming the inhomogeneous band and the ground-state occupancies of the ions. We do not calculate the background dielectric constant of the YSO crystal $\varepsilon_{\rm bkg}$, but instead assume the refractive index value found in the literature $n_{\rm bkg}=\sqrt{\varepsilon_{\rm bkg}}=1.8$ ~\cite{Beach1990}.

The validity of the Clausius-Mossotti equation depends on our ability to define regions within the solid where charge does not pass.  If such units do not exist~\cite{Martin1974}, then intrinsically solid-state approaches such as band theory~\cite{Levine1991} or the modern theory of polarizability~\cite{Resta1994} are required to calculate the dielectric function.
In the present case, the Pr$^{3+}$ impurity ions (a)~have open 4f shells which are located in the interior of the ions, (b)~are dilute, hence located far from each other on the scale of an atomic bond length, and (c)~individually have very narrow resonances, whose energies differ from ion to ion by many times the resonance width.  Each of these circumstances inhibits any conventional quantum mechanical interactions such as overlap of wave functions leading to bonding-antibonding splittings or band structure.
Hence, Martin's criterion~\cite{Martin1974} of a well-defined surface to separate the various oscillating charges is satisfied, building our confidence in the applicability of the Clausius-Mossotti equation.
The surface in question is simply a sphere drawn around each individual Pr$^{3+}$ ion well outside of the 4f electrons.

The sum over $k$ in Eq.~(\ref{eq:Clausius}) may be written as a sum over the
ground-state hyperfine levels $i$, integrated over the detuning $\Delta_0$ of their lowest-energy hyperfine transition $a$ (Fig.~\ref{fig:introFigure}). We measure $\Delta_0$ relative to the center of the inhomogeneous band $\omega_0$. We define
\begin{eqnarray}
B_i(\omega) = \frac{1}{\epsilon_0 \varepsilon_{\rm bkg}}\int \! \d\Delta_0 \,
\rho_i(\Delta_0) \, \alpha_i(\omega,\Delta_0)
\label{eq:BiA}
\end{eqnarray}
to rewrite Eq.~(\ref{eq:Clausius}) as
\begin{eqnarray}
3\frac{\varepsilon(\omega)-1}{\varepsilon(\omega)+2}
= \sum_i B_i(\omega)
\label{eq:sumBi}
\end{eqnarray}
where $\alpha_i(\omega,\Delta_0)$ is the polarizability of an ion for radiation at frequency $\omega$, assuming it is in ground-state hyperfine level $i$ and its lowest-energy transition is at frequency $\omega_0+\Delta_0$; and $\rho_i(\Delta_0)$ is the density of such ions. In our application, the frequency $\Delta_0$ runs over the inhomogeneous band, which is 4.4~GHz wide~\cite{Equall1995}, hence very narrow compared with the optical transition of 495~THz. The density $\rho_i$ includes both the assumed Gaussian background due to inhomogeneous broadening as well as any laser-induced redistribution of population, as occurs in AFC preparation. Redistribution preserves $\sum_i \rho_i(\Delta_0)$: spectral-hole burning can change the population distribution within the ground-state manifold but it cannot remove impurity ions or change their resonant frequencies. We assume there is no excited-state population. Experimentally, this is ensured by a 10~ms delay after hole burning --- about 61 times the mean time for spontaneous emission --- before the pulses generating the echoes begin.

In this work, $i$ runs over the three ground states $\gA$, $\gB$, and $\gC$, and we consider transitions to the three excited states $\eA$, $\eB$, and $\eC$. However, our formalism applies to arbitrary numbers of ground and excited states: this makes it readily adaptable to cases where the degenerate hyperfine levels are split due to external fields, and even to calculations for other dopant ions.

In the rotating-wave approximation (RWA), the atomic polarizability has the Drude-Lorentz form
\begin{eqnarray}
\alpha_i(\omega,\Delta_0) = f_0 \sum_j \frac{f_{ij}}{\omega_0+\Delta_0+\Delta^{(0)}_{ij} - \omega -\I\gamma}
\label{eq:alphaA}
\end{eqnarray}
where $j$ indexes the Pr$^{3+}$ ions' excited-state hyperfine levels, $\gamma$ is the inhomogeneous width, the $f_{ij}$ are the relative transition probabilities, and $f_0$ is a constant related to the dimensional oscillator strength. The $f_{ij}$ are constrained by conservation of probability $\sum_i f_{ij} = \sum_j f_{ij}=1$~\cite{Equall1995}. The constants $\Delta^{(0)}_{ij}$ are the transition frequencies relative to the $\gA\rightarrow\eA$ transition (Fig.~\ref{fig:introFigure}a). Note that the RWA leads to the near-resonance term in Eq.~(\ref{eq:alphaA}) as well as a term in the ultraviolet, which we have omitted: in our application, the neglected term is at least 5 orders of magnitude smaller than the term retained.

Ref.~\cite{Nilsson2004} gives oscillator strengths for the 9 hyperfine transitions in Pr:YSO. Within a manifold, the transition rates are proportional to these oscillator strengths. Oscillator strengths differ from squares of matrix elements by a factor proportional to the transition frequencies. This factor is $1+O(10^{-7})$ because the hyperfine splittings are less than 50~MHz and the reference transition frequency is 495~THz.

Eq.~(\ref{eq:alphaA}) allows us to write
\begin{eqnarray}
\alpha_i(\omega,\Delta_0) = \alpha_i(\omega-\Delta_0,0)
.\label{eq:alphaB}
\end{eqnarray}
Using the convention $\omega=\omega_0+\Delta$, the polarizability can be written as
\begin{eqnarray}
\tilde\alpha_i(\Delta) = f_0 \sum_j \frac{f_{ij}}{\Delta^{(0)}_{ij} - \Delta -\I\gamma}
.\label{eq:alphaC}
\end{eqnarray}
Similarly, we define $\tilde\varepsilon(\Delta)=\varepsilon(\omega)$ and $\tilde B_i(\Delta) = B_i(\omega)$. We may rewrite Eq.~(\ref{eq:BiA}) as
\begin{eqnarray}
\tilde B_i(\Delta) = \frac{1}{\epsilon_0}\int \! \d\Delta_0 \,
\rho_i(\Delta_0) \, \tilde\alpha_i(\Delta-\Delta_0)
\label{eq:BiB}
\end{eqnarray}
and Eq.~(\ref{eq:sumBi}) as
\begin{eqnarray}
3\frac{\tilde\varepsilon(\Delta)-1}{\tilde\varepsilon(\Delta)+2}
= \sum_i \tilde B_i(\Delta)
.\label{eq:sumBiTilde}
\end{eqnarray}
Eq.~(\ref{eq:BiB}) is a convolution integral.  As such, it can be evaluated rapidly by Fast Fourier transforms (FFTs), using the convolution theorem. A similar formulation was presented earlier~\cite{Sonajalg1994}.

To find the dielectric response of the unperturbed impurity band in our study, the three ground-state hyperfine levels were each assumed to have an identical Gaussian density of states with a full width at half maximum (FWHM) of 4400~MHz (standard deviation $\sigma_{\rm inh}= 1869$~MHz). When the crystal is unperturbed, all three states have equal occupancy. The frequency step is 50~kHz with $2^{20}$, i.e., over 1 million, points used in the FFT, leading to a bandwidth of 52~GHz, which is more than $28\,\sigma_{\rm inh}$. The polarization sums were formed on this grid leading to a function proportional to the polarizability, or equivalently $\varepsilon(\omega)-1$. The proportionality constant was set at the peak absorption as described in Section~\ref{subsec:propagation}.

\subsection{Pulse propagation}
\label{subsec:propagation}

Taking the beam direction in the crystal to be the $+z$ direction, a scalar wave propagating in such a medium obeys
\begin{eqnarray}
U(z) = U_0 \euler^{\I n k z} = U_0 \euler^{\I n_1 k z} \euler^{-n_2 k z}
\label{eq:scalarWave}
\end{eqnarray}
where $U_0$ is the incident optical amplitude, $k=2\pi n_{\rm bkg} / \lambda$ is the wavevector in an undoped YSO crystal, $n=\sqrt{\varepsilon}=n_1 + \I n_2$ is the adjustment to the refractive index $n_{\rm bkg}$ due to the dopant ions, and $\lambda$ is the free-space wavelength. The transmission is given by
\begin{eqnarray}
T(z) = \left| \frac{U(z)}{U_0}\right|^2 = \euler^{-2 n_2 k z}
.\label{eq:trans}
\end{eqnarray}
Assuming $T=9.9$~dB, as observed in our single-frequency-burn measurements in the 10~mm long crystal, we evaluate Eq.~(\ref{eq:trans}) at the center of the inhomogeneous band, finding $n_2(\omega_0)=1.1 \times 10^{-5}$. Here $n_1(\omega_0)=1$, leading to $\varepsilon(\omega_0)-1 = 2 \, \I \, n_2(\omega_0)$ where we neglect the term $[n_2(\omega_0)]^2$.

In practice, we find the sum of convolution integrals $\sum_i\tilde B_i(\Delta)$, divide by $\sum_i\tilde B_i(0)$ for the unperturbed system and multiply by $\tilde\varepsilon(0)-1$, found experimentally,
to obtain $\tilde\varepsilon(\Delta)-1$.
Thus, we avoid an explicit value for the density of impurity ions.

Our model of photon echoes consists of applying Eq.~(\ref{eq:scalarWave}) to each Fourier component of a pulse incident on the crystal with a waveform given by $U_0(t)$.  Because the pulse has a narrow bandwidth about a central frequency, the transform variable is the detuning $\Delta$.
Because the crystal has an antireflective (AR) coating, reflections are neglected. The dielectric response is given by $\tilde \varepsilon(\Delta)$.  At the exit face, the frequency components are back transformed to the time domain yielding $U_1(t)$ at the exit face.  Symbolically,
\begin{eqnarray}
U_1(t) = {\cal F}^{-1} \left\{ {\cal F} \left[ U_0(t) \right]
   \exp\left[ \I \tilde n(\Delta) k z \right]
                      \right\}
\label{eq:pulseFourier}
\end{eqnarray}
where $\tilde n(\Delta) = [\tilde\varepsilon(\Delta)]^{1/2}$
and $\cal F$ is the Fourier transform from time domain to the domain of detuned frequency $\Delta$.
A similar method was used by Teja et al.~\cite{teja2019photonic}.

\subsection{Calculations of ground-state population redistribution}
\label{subsec:denMat}

We calculate the AFC spectra, i.e., the redistribution of the ground-state population using two different methods, namely by solving the rate equations and by using a density matrix formalism. Each approach has strengths and weaknesses. The density matrix formalism is more generally applicable, while the rate equations may be derived from the density matrix under certain physical conditions.  When the rate equations are valid, only the populations, and not the coherences, need be calculated. In both cases, we treat each ion class independently: a more thorough analysis could consider the interplay between different classes, to include spectral diffusion effects, for example, but this is beyond the scope of our present work.

The rate equations, introduced by Einstein in 1905, have played a huge role in understanding optical systems such as lasers that depend on redistribution of population of quantum levels.  The density matrix is a more sophisticated level of theory which also includes quantum coherence.
While the rate equation captures the effects of absorption, stimulated emission, and spontaneous emission, it cannot capture coherent phenomena such as Rabi oscillations and coherent population redistribution mechanisms such as Stimulated Raman Adiabatic Passage (STIRAP)~\cite{Unanyan1998}.
Indeed, the name ``counterintuitive pulse sequence'' associated with STIRAP is ``counterintuitive'' under the unstated assumption that the intuition was built on the rate equations.

The text of Grynberg et al.~\cite{Grynberg2010} discusses the validity conditions for the rate equations.  Neither of the two conditions they discuss apply: neither do the coherences relax substantially faster than the populations are transferred, nor is the correlation time of the laser small compared to the decay time associated with the (power-broadened) absorption line.
Despite a question of the region of validity, we present results of atomic frequency combs based on the rate equations.  We also present a calculation of the redistribution of population using the density matrix.

We considered using the stochastic Schr{\"o}dinger equation (SSE)~\cite{Plenio1998} as well.  
An average of the SSE is formally equivalent to the density matrix~\cite{Johanson2012},
although the SSE has advantages for single-particle dynamics~\cite{Plenio1998}. Ref.~\cite{Johanson2012} suggests that the density matrix is favored algorithmically for small matrices, and the SSE is favored for large ones.  Ref.~\cite{Johanson2012} describes a cross-over when the dimension of the Hamiltonian exceeds 200.  Here, the Hamiltonian is only of dimension~6, suggesting the density matrix formalism will be more efficient. This is true, but it understates the case. Because our Hamiltonian is periodic, it is possible to obtain a solution in the logarithm of the number of periods.
Attal et al.~\cite{attal2018rate} make a similar observation to accelerate a similar calculation using the rate equation.
We were not able to find a comparable algorithm for the SSE, making the density matrix the better method for this application.

\subsubsection{Rate Equation Formalism}

In our rate equation calculations, we assume the entire population is initially distributed equally among the 3 ground-state levels, and describe their occupancies with a vector
\begin{equation}
N(t=0)=
\begin{pmatrix}
N_{1/2}(0) \\ N_{3/2}(0) \\ N_{5/2}(0)
\end{pmatrix}=
\begin{pmatrix}
1/3 \\ 1/3 \\ 1/3
\end{pmatrix}.
\label{eq:rateEqStart}
\end{equation}
Based on our linewidth measurements, we assign each ion class a Lorentzian absorption distribution around each of its transitions. To determine the amount of optical power addressing a given transition, we integrate the product of this distribution and the measured input optical spectrum used for hole burning. Repeating this process for all 9 transitions, we form a $3\times3$ matrix $I_{kj}$ of powers addressing the transitions from $|\,\mathbf{g},j\rangle$ to $|\,\mathbf{e},k\rangle$. We assume that the transition rates from $\mathbf{g}$ to $\mathbf{e}$ are proportional to these matrix elements, motivated by the fact that our sample has a high optical depth, so that there is a high probability of absorption regardless of the oscillator strength of a particular transition. Furthermore, we assume that the excited states decay to ground states within a time increment of our calculation, according to the oscillator strength matrix $p_{\ell k}$ describing transitions from $|\,\mathbf{e},k\rangle$ to $|\,\mathbf{g},\ell \rangle$ (the squares of the matrix elements in Table \ref{tab:amps}). The rate matrix $R_{\ell j}$ describing population redistribution from $|\,\mathbf{g},j\rangle$ to $|\,\mathbf{g},\ell\rangle$ has elements
\begin{equation}
    \begin{aligned}
    &R_{\ell j} = \sum_{k}p_{\ell k}I_{kj}, \;\;\;\ell\neq j\\
    &R_{jj} = - \! \sum_{\ell\neq j} R_{\ell j}
    \end{aligned}
    \label{eq:rateMatrixElements}
\end{equation}
Note that the off-diagonal elements are necessarily non-negative, since ions excited from $|\,\mathbf{g},j\rangle$ can only increase the population of $|\,\mathbf{g},\ell\neq j\rangle$; since these ions do not change the total population of all the ground states, the diagonal elements $R_{jj}\leq0$. The time evolution of the populations is calculated by
\begin{align}
    \Delta N_\ell(t)= \sum_j R_{\ell j}N_j(t)\Delta t,\\
    N(t+\Delta t)= N(t)+\Delta N(t)\nonumber.
    \label{eq:timeEvolution}
\end{align}

\subsubsection{Density Matrix Formulation}
The Lindblad master equation for the density matrix $\rho$ is
\begin{eqnarray}
\frac{\d\rho}{\d t} = -\I[H,\rho] +
\left. \frac{\d\rho}{\d t} \right|_{\rm relax},
\label{eq:LiouvilleRho}
\end{eqnarray}
where $H$ is the Hamiltonian and $\left. \frac{\d\rho}{\d t} \right|_{\rm relax}$ describes the relaxation.  We assume the relaxation is purely due to spontaneous emission and use the form given by Ref.~\cite{Grynberg2010}.

The further development of this equation requires an explicit form for the Hamiltonian.  Unfortunately, published Hamiltonian parameters are in conflict and are internally inconsistent.
Parameters for the Hamiltonian determined experimentally by Longdell et al.~\cite{Longdell2002} were later found to be in conflict with the empirical optical oscillator strengths of Nilsson et al.~\cite{Nilsson2004}. In an attempt to resolve this conflict, Lovri{\'c} et al.~\cite{Lovric2012} remeasured the Hamiltonian parameters. Unfortunately, these new parameters result in an oscillator-strength matrix with two columns swapped when compared with Nilsson et al. and even with their own work, as has been pointed out previously~\cite{Bartholomew2016}.

Since we were unable to find the Hamiltonian elsewhere in the literature, we derived a Hamiltonian that is consistent with the hyperfine level structure and oscillator strengths~\cite{Lovric2012} and that takes the conventional form $\frac{1}{2} I\cdot Q \cdot I$, where $I$ is a vector of length 3 representing the spin $\sfrac{5}{2}$ nuclear spin states and $Q$ is a rank-2 traceless symmetric tensor giving the orientation of these spins in a given manifold relative to the crystal axes. There are 10 degrees of freedom to be determined: the ground- and excited-state manifolds each have 2 energy spacings (Fig.~\ref{fig:introFigure}) and 3 Euler angles indicating their orientation with respect to the crystal axes. However, it is actually sufficient to know only the relative orientation of the manifolds, so we only need to determine 3 Euler angles. The 3$\times$3 oscillator-strength matrix given in Ref.~\cite{Equall1995} provides four constraints on these Euler angles. Using a least-squares fit we find Euler angles $\alpha_E=10.3(15)^\circ$,
$\beta_E = -164.4(15)^\circ$, and $\gamma_E=-130.7(15)^\circ$
(digits in parentheses, here and throughout, are total uncertainties at one standard error, and represent the uncertainty of the least significant digits).
The RMS deviation of the squares of the matrix elements in Table~\ref{tab:amps} from the oscillator strength values given in Ref.~\cite{Equall1995} is 0.013.  This value may be compared with the quadrature sum of the experimental uncertainty of 0.01 in Ref.~\cite{Equall1995} and the RMS uncertainty from the above fitting procedure, leading to a combined uncertainty of 0.023.

\begin{table}
\begin{center}
\caption{ 
Transition probability amplitudes between the $^3H_4$ and $^2D_1$ ($\mathbf{g}$ and $\mathbf{e}$) manifolds (the manifolds $\mathbf{g}$ and $\mathbf{e}$ are not to be confused with the transitions $g$ and $e$).
State labels are those of Fig.~1.
The signs, magnitudes, and reported uncertainties are found using the Euler angle fitting procedure described in the text.
}

\begin{tabular}{ cccc } 
 \hline
 \hline
 &  $\eA$ & $\eB$ & $\eC$ \\
\hline 
$\gA$ &
    $\phantom{-}0.753(23)$ & $-0.602(28)$ & $-0.265(24)$ \\
$\gB$ &
  $-0.634(27)$ & $-0.772(22)$ & $-0.048(08)$ \\
$\gC$ &
  $-0.176(17)$ &  $\phantom{-}0.204(19)$ & $-0.963(07)$ \\
 \hline
 \hline
\label{tab:amps}
\end{tabular}
\end{center}
\end{table}

Returning to the Lindblad Master equation, we proceed to a rotating frame
\begin{eqnarray}
\rho_{k\gamma,\ell\delta} = \sigma_{k\gamma,\ell\delta}
\euler^{-\I [\omega_0 + (\gamma-\delta) \Delta ]t}.
\label{eq:changeOfPicture}
\end{eqnarray}
Here, $\gamma$ and $\delta$ index the optical manifold (with 1 for $\mathbf{g}$ and 2 for $\mathbf{e}$), $k$ and $\ell$ index the hyperfine level within a given manifold,
and $\Delta$ is the detuning. Whereas $\rho$ varies with an optical frequency ($\approx$ 500~THz), $\sigma$ varies with the hyperfine splitting, a radio frequency (RF) of order 10~MHz, as shown in Fig.~\ref{fig:introFigure}a. This change of variables allows the optical frequencies to be removed analytically and allows the program to solve numerically on the RF time scale.
A time step of $\approx$ 1~ns is chosen, with a small adjustment often required to let the time step be commensurate with the period $T_{\rm rep}$.
The density matrix in the rotating frame is given by
\begin{eqnarray}
\frac{\d \sigma}{\d t} = {\cal L} \sigma
,\label{eq:LiouvilleSigma}
\end{eqnarray}
defining the Lindblad matrix $\cal L$ from context. This equation is solved by discretizing time and assuming that the Lindblad matrix is piecewise-constant in time.  In each time step, the solution is
\begin{eqnarray}
\sigma(t+\delta t) = \exp({\cal L} \delta t) \sigma(t)
.\label{eq:Propagator}
\end{eqnarray}
In practice, the matrix exponential is found through its Taylor series in the form used in Horner's method, namely
\begin{eqnarray}
\exp(M) =  1 + M (1 + \frac{M}{2}  ( 1 + \frac{M}{3} ( ... )))
\label{eq:MatrixExp}
\end{eqnarray}
for any matrix $M$. The terms are evaluated from the innermost parenthesis, starting from $\frac{M}{30}$.  This method proves to be more accurate than evaluating the eigenvalues and eigenvectors with the linear algebra package LAPACK~\cite{Lapack1999} and taking the exponential, stabilizing the solutions.

To evaluate the solution for a large number of frequency-modulated periods, we make use of the identity
\begin{eqnarray}
\int_0^{2T_{\rm rep}} \!\!\! \d t \; \exp[ {\cal L}(t) t] 
=
\left[ \int_0^{T_{\rm rep}} \!\!\! \d t \; \exp[ {\cal L}(t) t] \right]^2 ,
\label{eq:TwoPeriod}
\end{eqnarray}
where $T_{\rm rep}$ is the period of ${\cal L}(t)$. Iterating Eq.~(\ref{eq:TwoPeriod}) enables calculation for any integer multiple of the period $nT_{\rm rep}$ in a time proportional to $\log(n)$. Since the pulse duration is typically millions of times $T_{\rm rep}$, this yields an enormous savings in computation time compared to propagating in small time steps for the duration of the burn beam. The computation time to evaluate the propagator for a single period takes longer than its extension to millions of periods.

Finally, the Rabi overall coupling strength $\Omega_0$ is a fitting parameter.  We compare theory to experiment for the single-frequency burn given in Fig.~\ref{fig:singleFreqResponse}.  The linewidth is a function of $\Omega_0$, and the value that matches the FWHM of the central peak is shown in the plot. The five prominent transmission features are in good agreement with the experiment as well as the antiholes at positive detuning. The more complex pattern at negative detuning is in somewhat less satisfactory agreement with the experiment. For the AFCs, we present values for $\Omega_0=0.0125$. We also calculated for 8 other values between $\Omega_0=0.0032$ and $0.02$.  The results are quite similar in the range of 0.005 to~0.02, although the peak coefficient of correlation with the experimental results below is maximized with the value presented. 

The techniques developed in this section can be applied to periodic optical excitations of the crystal. In practice, the burn beam produced in the experiment is not exactly periodic, and the procedure for obtaining a periodic approximation to it is fairly subtle (Fig.~\ref{fig:threeSignal}). Details about this signal averaging step are in the Appendix.

In principle, spatial effects play a role in spectral-hole burning. For example, along the burn beam propagation direction, transmission peaks will become sharper as the beam is progressively filtered by the crystal.  We leave these refinements to future work.

To simulate our experimental AFC preparation, we run the density matrix code at 7001 frequencies over a bandwidth of 350~MHz (in 50~kHz steps which are smaller than the narrowest features experimentally observed when applying a single-frequency burn to the crystal, as shown in Fig.~\ref{fig:singleFreqResponse}), yielding predicted ground-state population fractions for each of three hyperfine levels. Outside of the 350~MHz band, the populations are assumed to be undisturbed. We multiply all population fractions by the Gaussian density of states given in Section~\ref{subsec:refIndex} to obtain the absorbing populations throughout the inhomogeneous band. The highly dispersive dielectric function of the AFC was calculated from these populations using Eq.~(\ref{eq:sumBiTilde}) and the equations leading to it in Section~\ref{subsec:refIndex}.

\begin{figure}
    \centering
    \includegraphics[width=15cm]{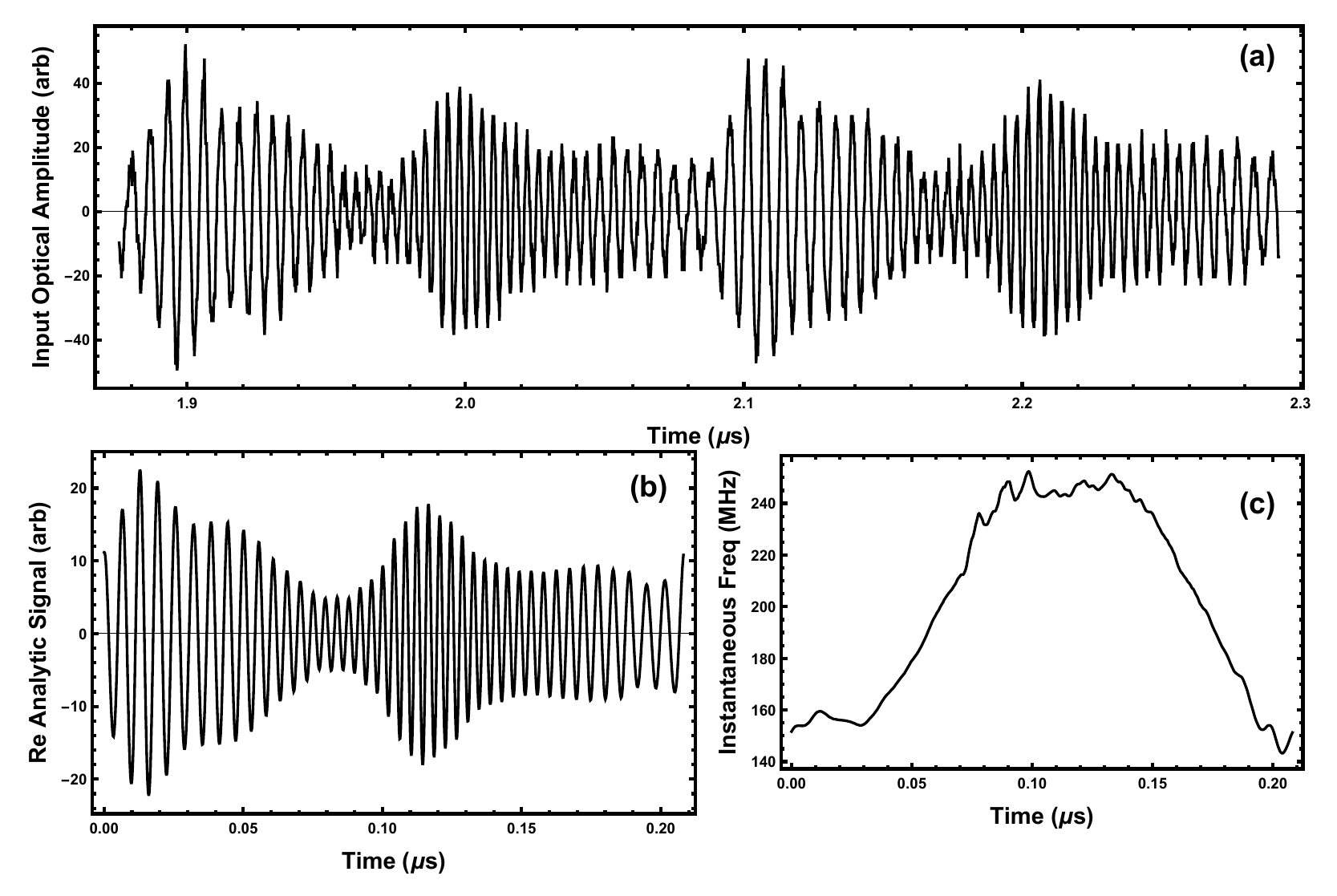}
    \caption{Input optical signals at 4.8~MHz optical-comb tooth spacing.  (a)~A portion of the output of the optical heterodyne signal showing the RF modulation of the optical input. (b)~Real part of analytic signal (Fourier transform with negative frequencies removed) obtained by a stroboscopic average of the magnitude of the analytic signal and, separately, the instantaneous frequency with a small frequency shift to ensure periodicity. (c)~Spectroscopic average of instantaneous frequency {vs.\@} time for the signal in part~(b).}
    \label{fig:threeSignal}
\end{figure}

\section{Results}
\label{sec:results}

\begin{figure}
  \centering
  \begin{center}
    \includegraphics[width=5in]{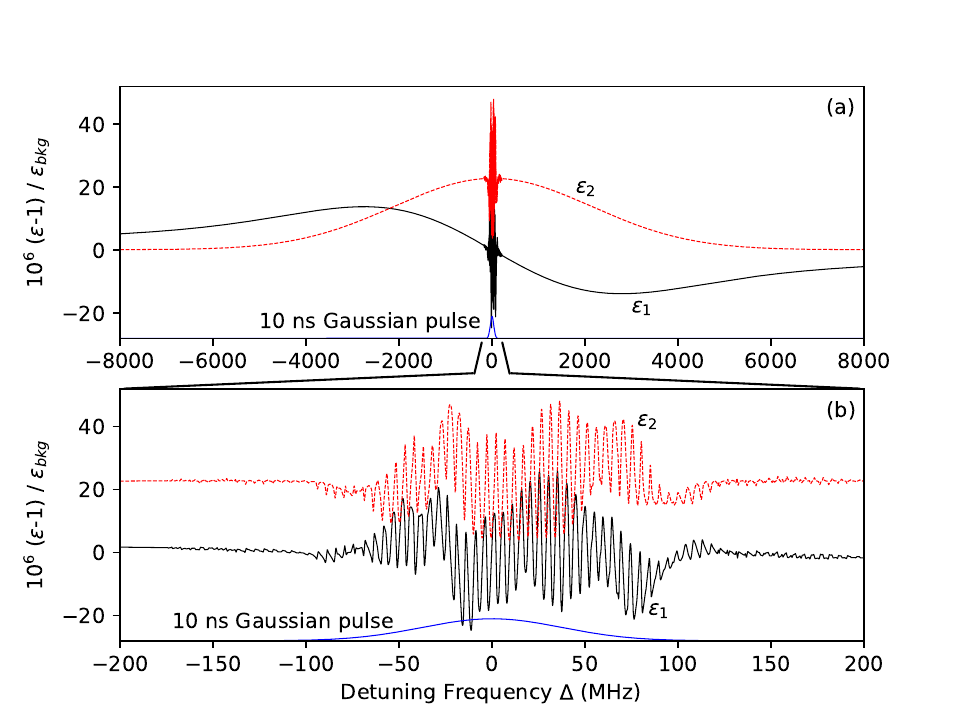}
  \end{center}
  \caption{
  Dielectric function calculated for the AFC with $f_{\rm rep}=4.9$~MHz with the probe pulse in frequency domain superimposed.
  (a) The dielectric function is given over the inhomogeneous band. (b) The dielectric function is given over the small portion of the inhomogeneous band which is modified due to the presence of the AFC.  The guide lines indicate that the lower graph is a magnified version of a portion of the upper graph.
  The real (solid black line) and imaginary (red dotted line) parts of the dielectric function are shown.   The specific function plotted is indicated on the $y$ axis.
  A 10~ns Gaussian pulse, representing the experimental pulse for photon echo generation (solid blue line) at the bottom of each panel is also shown.
  }
  \label{fig:dielectriFunction}
\end{figure}

\subsection{AFC transmission}
\label{subsec:transmissionResults}
A typical dielectric function calculated in the presence of an AFC is shown in Fig.~\ref{fig:dielectriFunction}. The AFC modifies only a small central region of the dielectric function: at large detunings, the AFC dielectric function approaches the unperturbed dielectric function. The key feature of the AFC dielectric function is its rapid oscillations with frequency.


\begin{figure}
 \centering
 \begin{center}
 \includegraphics[width=6in]{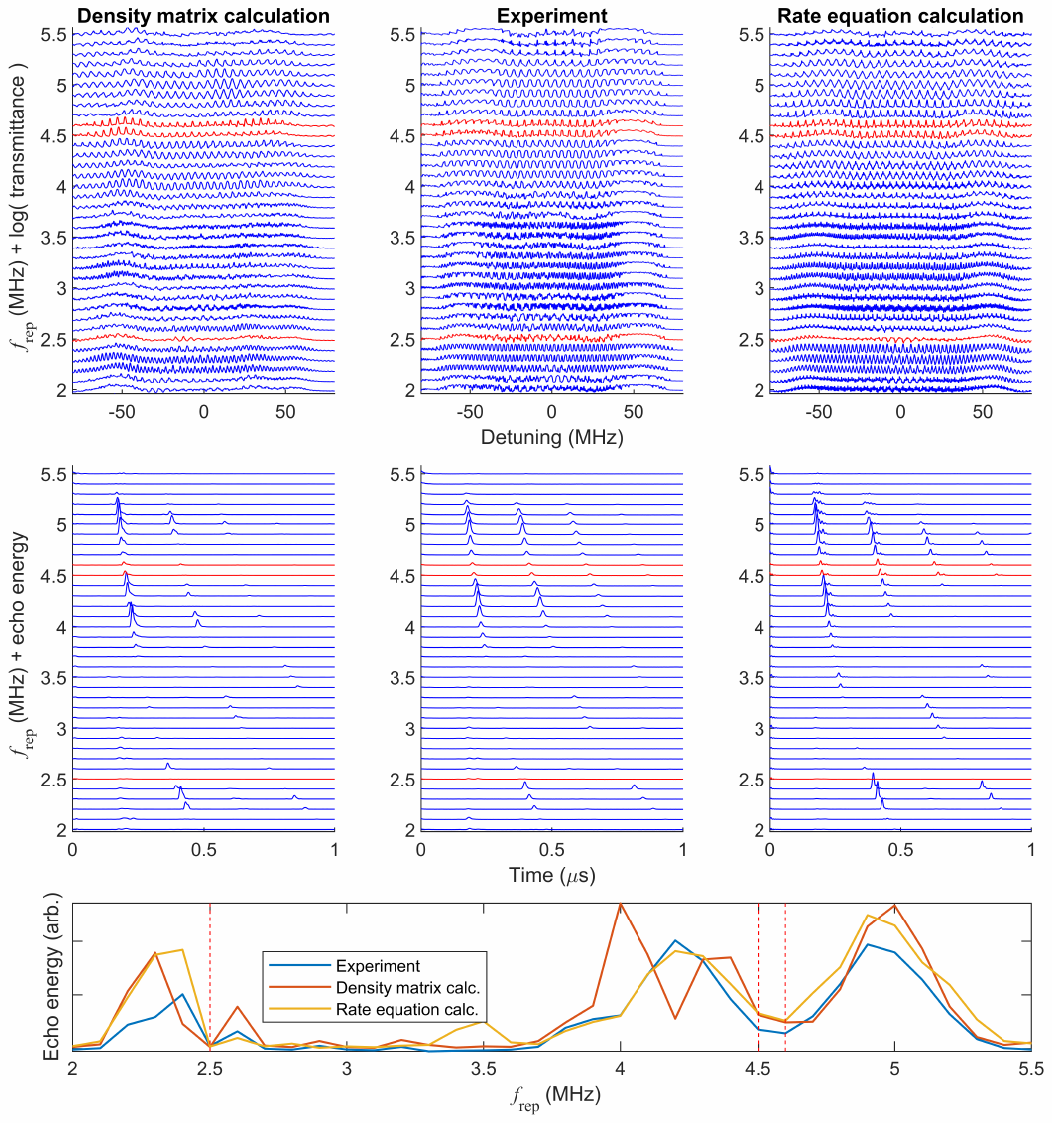}
 \end{center}
\caption*{FIG.~\ref{fig:7panelPlot}:  Caption on next page.
}
\end{figure}
\begin{figure}
\caption{\textit{(top 3)} Transmission spectra prepared using optical combs of varying tooth spacing $f_{\rm rep}$ from density matrix calculation, experiment, and rate-equation calculation; \textit{(center 3)} the corresponding echo pulses produced; \textit{(bottom)} the energy in the first-order echo pulse from density matrix (red), experiment (blue), and rate equation (yellow).
In the first two rows, the vertical offset of each curve indicates $f_{\rm rep}$; the curves highlighted in red indicate local minima of the experimental first-order echo energy with respect to $f_{\rm rep}$. These minima are indicated by vertical dashed lines in the bottom plot.}
 \label{fig:7panelPlot}
\end{figure}

The transmission spectra can be calculated from the dielectric function using Eq.~(\ref{eq:trans}). Fig.~\ref{fig:7panelPlot} shows experimental results along with the corresponding theoretical spectra for both the rate equation and density matrix. The theories capture many features of the experimental transmission spectra, such as the widths and shapes of transmission windows between comb teeth, as well as transitions to non-periodic forms.  The rate equations tend to predict more regular combs than the density matrix theory. We see that combs with large fundamental Fourier components occur a few hundred kHz above and below 4.7~MHz, as opposed to at 4.7~MHz where the combs are not dominated by the fundamental~\cite{Weiner1990a,Slattery2021}. The effect of these features is discussed further in Section~\ref{subsec:echoResults}.


\begin{figure}
 \centering
  \begin{center}
 \includegraphics[width=6.3in]{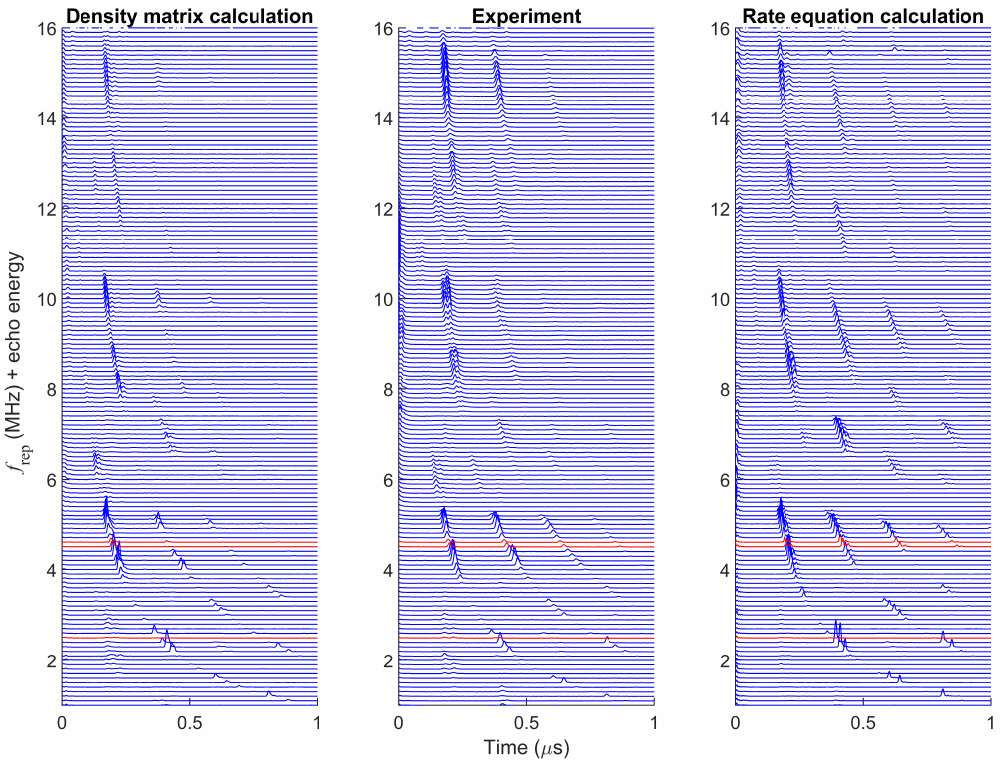}
  \end{center}
 \caption{Echo pulses from the density matrix calculation, experiment, and rate equation calculation, for $f_{rep}$ ranging from 1~MHz to 16~MHz, in steps of 0.1~MHz. The lines are offset vertically by the repetition rate.}
 \label{fig:pulse}
\end{figure}

\subsection{Photon echoes}
\label{subsec:echoResults}

The input pulse was modeled as a Gaussian in the time domain with 10~ns FWHM, corresponding to 88~MHz FWHM in the frequency domain. The pulse bandwidth is compared with the AFC bandwidth in Fig.~\ref{fig:dielectriFunction}, showing that it is a good match for the oscillatory region of the AFC dielectric function. After discarding spurious points arising from the artificial periodicity of the FFT, we obtain a time span of 10~$\mu$s, several times longer than the 1~$\mu$s time period in which the pulses appear. Fig.~\ref{fig:7panelPlot} shows the experimental AFCs (top center) alongside density matrix (top left) and rate equation (top right) calculations, and the corresponding pulses (bottom). Fig.~\ref{fig:pulse} shows the echo pulses over a wider range of $f_{\rm rep}$ values.
Although Fig.~\ref{fig:dielectriFunction} is given for one of the density matrix calculations at a single repetition rate, we find the dielectric function separately for each theory at each repetition rate.

The initial pulse near 0~$\mu$s is present in both experiment and theory. The echo pulses appear on hyperbolas, with the $N^{\rm th}$-order echo emission occurring at $T=N/f_{\rm rep}$. Fig.~\ref{fig:7panelPlot} highlights a  number of features common to the experiment and the theory. Most notably, $f_{\rm rep}$ the largest first-order echoes do not align well with the excited-state  hyperfine energy spacing values (dashed lines in panel 5): the largest echoes occur for $f_{\rm rep}$ detuned slightly above and below these energy spacings. Indeed, the echo efficiency at one of those spacings is near its lowest value.  The patterns of drop-out of the first and second photon echoes is predicted well: for example, the absence of a photon echo for $f_{\rm rep}=2.5$~MHz is predicted by both theories and it is experimentally observed, despite photon echoes for both $f_{\rm rep}=2.4$~MHz and 2.6~MHz.
Also common to both theories and experiment are valleys near 3.5~MHz and 4.5~MHz,  although the drop-out is over-predicted by theory for the 4.5 MHz case. There are also small echoes in both experiment and density matrix theory for $f_{\rm rep}$ between 2.0 MHz and 3.5 MHz, at nearly fixed delay times between 0.20~$\mu$s and 0.25~$\mu$s.  These are much less prominent in the rate equation calculation.

\section{Conclusions}
\label{sec:conclusion}
We present a semiclassical theory of photon echoes that gives a good account of the pulses arising in an AFC experiment in Pr:YSO with fine-grained systematic spectral coverage.  The theory has two variants, one in which the AFC is predicted by the Rate Equations and a second in which the AFC is predicted by a Density Matrix.
In contrast to most descriptions of AFC's in the literature, our approach does not invoke the Dicke state~\cite{Dicke1954}. The theory involves obtaining the dielectric function of the contribution of the Pr$^{3+}$ ions to the crystal from (a) a rate-equation or density-matrix description of the relevant hyperfine levels in the Pr$^{3+}$ impurity ions leading to the atomic polarizability (up to a constant) along with the input optical signals that prepared the AFC, (b) the Clausius-Mossotti relation, and (c) a fit to the maximum absorption in the case of a crystal without an AFC (thus fixing the constant).
The photon echoes are modeled by letting the probe pulse pass through the AFC, here modeled as a highly dispersive medium. There does not appear to have been any similar attempt to simulate realistic atomic frequency combs based on the actual optical fields used to prepare them. Further, by generating over 100 AFCs for our systematic study, we demonstrate the predictive power of our theoretical model.

Looking ahead, having a predictive model will enable tailoring the AFC preparation protocol to optimize the efficiency of the system as a quantum memory.  The fact that the predictive model is relatively simple in that it relies on long-established principles of the response of an oscillator to classical electromagnetic fields will hopefully enable applications.



\section{Acknowledgments}
We are grateful to Alexey Gorshkov, Robinjeet Singh, and Eite Tiesinga for helpful discussions and to Ivan Burenkov, and Sergey Polyakov for helping to build the experimental setup.
Mention of commercial products does not imply endorsement by the authors' institutions.

\section*{Appendix}
\subsection*{Signal averaging}
The theory requires knowledge of the periodic signal entering the crystal.  Here, we describe how that periodic signal is obtained from the RF recorded after heterodyne mixing, shown in Fig.~\ref{fig:expLayout}a.  Signal averaging is used to extract the periodic signal from 20~$\mu$s samples of data discretized in 0.2~ns steps. The algorithm for averaging the signals is given below.

Results for the case of a modulation frequency $f_{\rm rep} = 4.8$~MHz are shown in Fig.~\ref{fig:threeSignal}, including the data before and after the averaging procedure.  The instantaneous frequency shown in Fig.~\ref{fig:threeSignal}c may be compared to the intended waveform, an ideal triangle-wave modulation: the discrepancy is due to limitations of the wave form generator.

The quality of the averaged signal may be assessed by comparing its Fourier transform to the Fourier transform of the input signal, as shown for the example of 4.8~MHz in Fig.~\ref{fig:fourierComparison}. The averaged spectrum is necessarily discrete. The experimental spectrum shows strong peaks that are broadened in analogy with x-ray-diffraction Bragg peaks broadened through thermal motions of the atoms, as described by the Debye-Waller factor.
The alignment of the discrete points to the peaks of the black curve validates the averaging process.  Matches of similar quality were obtained for each of the 161 modulation frequencies from 1~MHz to 16~MHz in steps of 0.1~MHz. The signal averaging algorithm has the following steps:
\begin{enumerate}
    \item  The mean is subtracted from the sample and the sample is smoothed with a 7-point symmetric binomial filter.  
    \item Locations of zero crossings (i.e., sign switches in adjacent smoothed data points) are found. If there are two zero crossings in consecutive time steps, the pair is not considered in the next step.
    \item The instantaneous frequency $f\approx \frac{1}{2\pi} \frac{\Delta \phi}{\Delta t}$ is calculated by assigning a phase shift of $\Delta \phi =\pi$ to adjacent zero crossings which also give the time interval $\Delta t$.
    \item The instantaneous frequency is averaged stroboscopically given the known driving period.
    \item The envelope is obtained by extracting the analytic signal from the smoothed data of step 2 above. The analytic signal is obtained through discrete FFT followed by zeroing out the negative frequencies, followed by the inverse FFT.  The envelope is the absolute value of the analytic signal.
    \item The envelope is averaged stroboscopically to create a periodic function.
    \item The periodic phase is obtained by integrating the averaged instantaneous frequencies obtained in step~3 over a period.
    A small linear adjustment in the phase is made to enforce strict periodicity of both amplitude and phase.
    \item The signal, a complex quantity, is formed as the product of the envelope from step~6 and the phasor of the periodic phase from step~7. (The phasor of a phase $\phi$ is $\euler^{\I \phi}$.)
    \item The signal is shifted in the Fourier domain to remove the central modulating frequency.  This corresponds physically to incorporating the central frequency of the AOM into the optical reference signal.  The effect is to create a function that may be sampled with more widely spaced points.  In practice, there is a 5:1 downsampling at this step from 0.2~ns to near 1~ns.  The step sizes are not exactly equal for different repetition rates because each time step must be an integer divisor of the corresponding period.
\end{enumerate}

\bibliographystyle{ieeetr}
\bibliography{bibliography}

\begin{thebibliography}{10}

\bibitem{Acin2018}
A.~Ac{\'\i}n, I.~Bloch, H.~Buhrman, T.~Calarco, C.~Eichler, J.~Eisert,
  D.~Esteve, N.~Gisin, S.~J. Glaser, F.~Jelezko, S.~Kuhr, M.~Lewenstein, M.~F.
  Riedel, P.~O. Schmidt, R.~Thew, A.~Wallraff, I.~Walmsley, and F.~K. Wilhelm,
  ``The quantum technologies roadmap: a {European} community view,'' {\em New
  J. Phys.}, vol.~20, p.~080201, 2018.

\bibitem{Pompili2021}
M.~Pompili, S.~L.~N. Hermans, S.~Baier, H.~K.~C. Beukers, P.~C. Humphreys,
  R.~N. Schouten, R.~F.~L. Vermeulen, M.~J. Tiggelman, L.~{dos Santos Martins},
  B.~Dirkse, S.~Wehner, and R.~Hanson, ``Realization of a multinode quantum
  network of remote solid-state qubits,'' {\em Science}, vol.~372,
  p.~259–264, 2021.

\bibitem{Briegel1998}
H.-J. Briegel, W.~D\"ur, J.~I. Cirac, and P.~Zoller, ``Quantum repeaters: The
  role of imperfect local operations in quantum communication,'' {\em Phys.
  Rev. Lett.}, vol.~81, pp.~5932--5935, Dec 1998.

\bibitem{Duan2001}
L.-M. Duan, M.~D. Lukin, J.~I. Cirac, and P.~Zoller, ``Long-distance quantum
  communication with atomic ensembles and linear optics,'' {\em Nature},
  vol.~414, pp.~413--418, 2001.

\bibitem{Awschalom2018}
D.~D. Awschalom, R.~Hanson, J.~Wrachtrup, and B.~B. Zhou, ``{Quantum
  technologies with optically interfaced solid-state spins},'' {\em {Nature
  Phot.}}, vol.~{12}, pp.~{516--527}, {2018}.

\bibitem{simon2007quantum}
C.~Simon, H.~De~Riedmatten, M.~Afzelius, N.~Sangouard, H.~Zbinden, and
  N.~Gisin, ``Quantum repeaters with photon pair sources and multimode
  memories,'' {\em Physical review letters}, vol.~98, no.~19, p.~190503, 2007.

\bibitem{Afzelius2009}
M.~Afzelius, C.~Simon, H.~de~Riedmatten, and N.~Gisin, ``Multimode quantum
  memory based on atomic frequency combs,'' {\em Phys. Rev. A}, vol.~79,
  p.~052329, May 2009.

\bibitem{Afzelius2010}
M.~Afzelius, I.~Usmani, A.~Amari, B.~Lauritzen, A.~Walther, C.~Simon,
  N.~Sangouard, J.~Min{\'a}{\v r}, H.~de~Riedmatten, N.~Gisin, and
  S.~Kr{\"o}ll, ``Demonstration of atomic frequency comb memory for light with
  spin-wave storage,'' {\em Phys. Rev. Lett.}, vol.~104, p.~040503, 2010.

\bibitem{Graf1997}
F.~R. Graf, A.~Renn, and U.~P. Wild, ``Site interference in stark-modulated
  photon echoes,'' {\em Phys. Rev. B}, vol.~55, pp.~11225--11229, 1997.

\bibitem{Graf1998}
F.~R. Graf, A.~Renn, G.~Zumofen, and U.~P. Wild, ``Photon-echo attenuation by
  dynamical processes in rare-earth-ion doped crystals,'' {\em Phys. Rev. B},
  vol.~58, pp.~5462--5478, 1998.

\bibitem{Nilsson2004}
M.~Nilsson, L.~Rippe, S.~Kr\"oll, R.~Klieber, and D.~Suter, ``Hole-burning
  techniques for isolation and study of individual hyperfine transitions in
  inhomogeneously broadened solids demonstrated in
  {Pr}$^{3+}$:{Y}$_2${SiO}$_5$,'' {\em Phys. Rev. B}, vol.~70, p.~214116, 2004.

\bibitem{beavan2013demonstration}
S.~E. Beavan, E.~A. Goldschmidt, and M.~J. Sellars, ``Demonstration of a
  dynamic bandpass frequency filter in a rare-earth ion-doped crystal,'' {\em
  JOSA B}, vol.~30, no.~5, pp.~1173--1177, 2013.

\bibitem{Fan2019}
H.~Q. Fan, K.~H. Kagalwala, S.~V. Polyakov, A.~L. Migdall, and E.~A.
  Goldschmidt, ``Electromagnetically induced transparency in inhomogeneously
  broadened solid media,'' {\em Phys. Rev. A}, vol.~99, p.~053821, 2019.

\bibitem{Klein2007}
J.~Klein, F.~Beil, and T.~Halfmann, ``Robust population transfer by stimulated
  {R}aman adiabatic passage in a {Pr}$^{3+}$:{Y}$_2${SiO}$_5$ crystal,'' {\em
  Phys. Rev. Lett.}, vol.~99, p.~113003, 2007.

\bibitem{Gao2007}
H.~Goto and K.~Ichimura, ``Observation of coherent population transfer in a
  four-level tripod system with a rare-earth-metal-doped crystal,'' {\em Phys.
  Rev. A}, vol.~75, p.~033404, 2007.

\bibitem{Wang2008}
H.-H. Wang, L.~Wang, X.-G. Wei, Y.-J. Li, D.-M. Du, Z.-H. Kang, Y.~Jiang, and
  J.-Y. Gao, ``Storage and selective release of optical information based on
  fractional stimulated {R}aman adiabatic passage in a solid,'' {\em Appl.
  Phys. Lett.}, vol.~92, p.~041107, 2008.

\bibitem{Mieth2012}
S.~Mieth, D.~Schraft, T.~Halfmann, and L.~P. Yatsenko, ``Rephasing of optically
  driven coherences by rapid adiabatic passage in
  {Pr}$^{3+}$:{Y}$_2${SiO}$_5$,'' {\em Phys. Rev. A}, vol.~86, p.~063404, 2012.

\bibitem{Goldschmidt2013}
E.~A. Goldschmidt, S.~E. Beavan, S.~V. Polyakov, A.~L. Migdall, and M.~J.
  Sellars, ``Storage and retrieval of collective excitations on a long-lived
  spin transition in a rare-earth ion-doped crystal,'' {\em Opt. Express},
  vol.~21, pp.~10087--10094, 2013.

\bibitem{Longdell2005}
J.~J. Longdell, E.~Fraval, M.~J. Sellars, and N.~B. Manson, ``Stopped light
  with storage times greater than one second using electromagnetically induced
  transparency in a solid,'' {\em Phys. Rev. Lett.}, vol.~95, no.~6, p.~063601,
  2005.

\bibitem{Heinze2013}
G.~Heinze, C.~Hubrich, and T.~Halfmann, ``Stopped light and image storage by
  electromagnetically induced transparency up to the regime of one minute,''
  {\em Phys. Rev. Lett.}, vol.~111, p.~033601, 2013.

\bibitem{Kutluer2016}
K.~Kutluer, M.~F. Pascual-Winter, J.~Dajczgewand, P.~M. Ledingham, M.~Mazzera,
  T.~Chaneli{\`e}re, and H.~de~Riedmatten, ``Spectral-hole memory for light at
  the single-photon level,'' {\em Phys. Rev. A}, vol.~93, p.~040302(R), 2016.

\bibitem{Hedges2010}
M.~P. Hedges, J.~J. Longdell, Y.~Li, and M.~J. Sellars, ``Efficient quantum
  memory for light,'' {\em Nature}, vol.~465, no.~7301, pp.~1052--1056, 2010.

\bibitem{Rielander2014}
D.~Riel{\"a}nder, K.~Kutluer, P.~M. Ledingham, M.~G{\"u}ndo{\v g}an, J.~Fekete,
  M.~Mazzera, and H.~de~Riedmatten, ``Quantum storage of heralded single
  photons in a praseodymium-doped crystal,'' {\em Phys. Rev. Lett.}, vol.~112,
  p.~040504, 2014.

\bibitem{Seri2019}
A.~Seri, D.~Lago-Rivera, A.~Lenhard, G.~Corrielli, R.~Osellame, M.~Mazzera, and
  H.~de~Riedmatten, ``Quantum storage of frequency-multiplexed heralded single
  photons,'' {\em Phys. Rev. Lett.}, vol.~123, no.~8, p.~080502, 2019.

\bibitem{Mannami2021}
K.~Mannami, T.~Kondo, T.~Tsuno, T.~Miyashita, D.~Yoshida, K.~Ito, K.~Niizeki,
  I.~Nakamura, F.-L. Hong, and T.~Horikiri, ``Coupling of a quantum memory and
  telecommunication wavelength photons for high-rate entanglement distribution
  in quantum repeaters,'' {\em Opt. Express}, vol.~25, pp.~41522--41533, 2021.

\bibitem{Horvath2021}
S.~P. Horvath, M.~K. Alqedra, A.~Kinos, A.~Walther, J.~M. Dahlstr{\"o}m,
  S.~Kr{\"o}ll, and L.~Rippe, ``Noise-free on-demand atomic frequency comb
  quantum memory,'' {\em Phys. Rev. Research}, vol.~3, p.~023099, 2021.

\bibitem{Bonarota2010}
M.~Bonarota, J.~Ruggiero, J.~L.~L. Gou\"et, and T.~Chaneli\`ere, ``Efficiency
  optimization for atomic frequency comb storage,'' {\em Phys. Rev. A},
  vol.~81, p.~033803, Mar 2010.

\bibitem{Usmani2010}
I.~Usmani, M.~Afzelius, H.~De~Riedmatten, and N.~Gisin, ``Mapping multiple
  photonic qubits into and out of one solid-state atomic ensemble,'' {\em
  Nature Commun.}, vol.~1, no.~1, pp.~1--7, 2010.

\bibitem{Davidson2020}
J.~H. Davidson, P.~Lefebvre, J.~Zhang, D.~Oblak, and W.~Tittel, ``Improved
  light-matter interaction for storage of quantum states of light in a
  thulium-doped crystal cavity,'' {\em Phys. Rev. A}, vol.~{101}, p.~042333,
  {2020}.

\bibitem{Saglamyurek2011}
E.~Saglamyurek, N.~Sinclair, J.~Jin, J.~A. Slater, D.~Oblak, F.~Bussieres,
  M.~George, R.~Ricken, W.~Sohler, and W.~Tittel, ``Broadband waveguide quantum
  memory for entangled photons,'' {\em Nature}, vol.~469, no.~7331,
  pp.~512--515, 2011.

\bibitem{Jobez2016}
P.~Jobez, N.~Timoney, C.~Laplane, J.~Etesse, A.~Ferrier, P.~Goldner, N.~Gisin,
  and M.~Afzelius, ``Towards highly multimode optical quantum memory for
  quantum repeaters,'' {\em Phys. Rev. A}, vol.~93, p.~032327, 2016.

\bibitem{Zhong2017}
T.~Zhong, M.~Kindem, J.~G. Bartholomew, J.~Rochman, I.~Craiciu, E.~Miyazono,
  M.~Bettinelli, E.~Cavalli, V.~Verma, S.~W. Nam, F.~Marsili, M.~D. Shaw, A.~D.
  Beyer, and A.~Faraon, ``Nanophotonic rare-earth quantum memory with optically
  controlled retrieval,'' {\em Science}, vol.~357, pp.~1392--1395, 2017.

\bibitem{Askarani2019}
M.~F. Askarani, M.~G. Puigibert, T.~Lutz, V.~B. Verma, M.~D. Shaw, S.~W. Nam,
  N.~Sinclair, D.~Oblak, and W.~Tittel, ``Storage and reemission of heralded
  telecommunication-wavelength photons using a crystal waveguide,'' {\em Phys.
  Rev. Applied}, vol.~11, p.~054056, May 2019.

\bibitem{Businger2020}
M.~Businger, A.~Tiranov, K.~T. Kaczmarek, S.~Welinski, Z.~Zhang, A.~Ferrier,
  P.~Goldner, and M.~Afzelius, ``Optical spin-wave storage in a solid-state
  hybridized electron-nuclear spin ensemble,'' {\em {Phys. Rev. Lett.}},
  vol.~{124}, p.~{053606}, {2020}.

\bibitem{Saglamyurek2016}
E.~Saglamyurek, M.~G. Pugibert, Q.~Zhao, L.~Giner, F.~Marsili, V.~B. Verma,
  S.~W. Nam, L.~Oesterling, D.~Nippa, D.~Oblak, and W.~Tittel, ``A multiplexed
  light-matter interface for fibre-based quantum networks,'' {\em Nature
  Commun.}, vol.~7, p.~11202, 2016.

\bibitem{Craiciu2021}
I.~Craiciu, M.~Lei, J.~Rochman, J.~G. Bartholomew, and A.~Faraon,
  ``Multifunctional on-chip storage at telecommunication wavelength for quantum
  networks,'' {\em Optica}, vol.~8, pp.~114--121, 2021.

\bibitem{Sabooni2013}
M.~Sabooni, Q.~Li, S.~Kr\"oll, and L.~Rippe, ``Efficient quantum memory using a
  weakly absorbing sample,'' {\em Phys. Rev. Lett.}, vol.~110, p.~133604, 2013.

\bibitem{Jobez2014}
P.~Jobez, I.~Usmani, N.~Timoney, C.~Laplane, N.~Gisin, and M.~Afzelius,
  ``Cavity-enhanced storage in an optical spin-wave memory,'' {\em New J.
  Phys.}, vol.~16, no.~8, p.~083005, 2014.

\bibitem{Nicolle2021}
M.~Nicole, J.~N. Becker, C.~Wientezl, I.~A. Walmsley, and P.~M. Leddingham,
  ``Gigahertz bandwidth optical memory in {Pr$^{3+}$:YSO},'' {\em arXiv},
  vol.~2102.13113v1, 2021.

\bibitem{teja2019photonic}
G.~P. Teja, C.~Simon, and S.~K. Goyal, ``Photonic quantum memory using an
  intra-atomic frequency comb,'' {\em Phys. Rev. A}, vol.~99, p.~052314, May
  2019.

\bibitem{Dicke1954}
R.~H. Dicke, ``Coherence in spontaneous radiation processes,'' {\em Phys.
  Rev.}, vol.~93, pp.~99--110, 1954.

\bibitem{Sangouard2007}
N.~Sangouard, C.~Simon, M.~Afzelius, and N.~Gisin, ``Analysis of a quantum
  memory for photons based on controlled reversible inhomogeneous broadening,''
  {\em Phys. Rev. A}, vol.~75, p.~032327, 2007.

\bibitem{Gorshkov2007b}
A.~V. Gorshkov, A.~Andr{\'e}, M.~D. Lukin, and A.~S. S{\o}rensen, ``Photon
  storage in $\lambda$-type optically dense atomic media. ii. free-space
  model,'' {\em Phys. Rev. A}, vol.~76, p.~033805, 2007.

\bibitem{Chaneliere2018}
T.~Chaneli{\`e}re, H.~H{\'e}tet, and N.~Sangouard, ``Quantum optical memory
  protocols in atomic ensembles,'' {\em Adv. Atom. Molec. and Opt. Phys.},
  vol.~67, pp.~77--150, 2018.

\bibitem{thiel2014measuring}
C.~Thiel, R.~Macfarlane, Y.~Sun, T.~B{\"o}ttger, N.~Sinclair, W.~Tittel, and
  R.~Cone, ``Measuring and analyzing excitation-induced decoherence in
  rare-earth-doped optical materials,'' {\em Laser Physics}, vol.~24, no.~10,
  p.~106002, 2014.

\bibitem{xiong2008numerical}
J.~Xiong, M.~Colice, F.~Schlottau, K.~Wagner, and B.~Fornberg, ``Numerical
  solutions to 2d maxwell--bloch equations,'' {\em Optical and quantum
  electronics}, vol.~40, no.~5, pp.~447--453, 2008.

\bibitem{arslanov2017optimal}
N.~M. Arslanov and S.~A. Moiseev, ``Optimal periodic frequency combs for
  high-efficiency optical quantum memory based on rare-earth ion crystals,''
  {\em Quantum Electronics}, vol.~47, no.~9, p.~783, 2017.

\bibitem{tittel2010photon}
W.~Tittel, M.~Afzelius, T.~Chaneliere, R.~L. Cone, S.~Kr{\"o}ll, S.~A. Moiseev,
  and M.~Sellars, ``Photon-echo quantum memory in solid state systems,'' {\em
  Laser \& Photonics Reviews}, vol.~4, no.~2, pp.~244--267, 2010.

\bibitem{Maksimov1969crystal}
B.~Maksimov, Y.~A. Kharitonov, V.~Ilyukhin, and N.~Belov, ``Crystal structure
  of the {Y-Oxysilicate Y [SiO 4] O},'' {\em Sov. Phys. Doklady}, vol.~13,
  p.~1188, 1969.

\bibitem{Holliday1993}
K.~Holliday, M.~Croci, E.~Vauthey, and U.~P. Wild, ``Spectral hole burning and
  holography in an {Y}$_2${SiO}$_5$:{Pr}$^{3+}$ crystal,'' {\em Phys. Rev. B},
  vol.~47, pp.~14741--14752, 1993.

\bibitem{Equall1995}
R.~W. Equall, R.~L. Cone, and R.~M. Macfarlane, ``Homogeneous broadening and
  hyperfine structure of optical transitions in {Pr}$^{3+}$:{Y}$_2${SiO}$_5$,''
  {\em Phys. Rev. B}, vol.~52, pp.~3963--3969, 1995.
\newblock In order to enforce the $f$-sum rule exactly, we make the following
  small adjustments $0.55\rightarrow0.552$, $0.38\rightarrow0.382$,
  $0.07\rightarrow0.066$, $0.40\rightarrow0.396$, $0.60\rightarrow0.596$,
  $0.01\rightarrow0.008$, $0.05\rightarrow0.052$, $0.02\rightarrow0.022$, and
  $0.93\rightarrow0.926$. These corrections are smaller than the measurement
  uncertainty which is 0.01.

\bibitem{Black2001}
E.~D. Black, ``An introduction to {Pound-Drever-Hall} laser frequency
  stabilization,'' {\em Am. J. Phys.}, vol.~69, pp.~69--87, 2001.

\bibitem{note:cycleNumber}
We have automated the experiment: due to variations in the computer speed over
  the course of the hours-long measurements, there is a small variation in the
  number of measurements for each modulation frequency.

\bibitem{Minar2010}
J.~Min{\'a}{\v r}, N.~Sangouard, M.~Afzelius, H.~de~Riedmatten, and N.~Gisin,
  ``Spin-wave storage using chirped control fields in atomic frequency
  comb-based quantum memory,'' {\em Phys. Rev. A}, vol.~82, p.~042309, 2010.

\bibitem{Jin2015}
J.~Jin, E.~Saglamyurek, M.~l.~G. Puigibert, V.~Verma, F.~Marsili, S.~W. Nam,
  D.~Oblak, and W.~Tittel, ``Telecom-wavelength atomic quantum memory in
  optical fiber for heralded polarization qubits,'' {\em Phys. Rev. Lett.},
  vol.~115, p.~140501, Sep 2015.

\bibitem{Furuya2020}
K.~Furuya, A.~Nandi, and M.~Hosseini, ``Study of atomic geometry and its effect
  on photon generation and storage [invited],'' {\em Opt. Mat. Express},
  vol.~10, pp.~577--587, 2020.

\bibitem{Holzapfel2020}
A.~Holzapfel, J.~Etesse, K.~T. Kaczmarek, A.~Tiranov, N.~Gisin, and
  M.~Afzelius, ``Optical storage for 0.53 s in a solid-state atomic frequency
  comb memory using dynamical decoupling,'' {\em New J. Phys.}, vol.~22,
  p.~063009, 2020.

\bibitem{Etesse2021}
J.~Etesse, A.~Holz\"apfel, A.~Ortu, and M.~Afzelius, ``Optical and spin
  manipulation of non-kramers rare-earth ions in a weak magnetic field for
  quantum memory applications,'' {\em Phys. Rev. A}, vol.~103, p.~022618, Feb
  2021.

\bibitem{chaneliere2010efficient}
T.~Chaneli{\`e}re, J.~Ruggiero, M.~Bonarota, M.~Afzelius, and J.~LeGouet,
  ``Efficient light storage in a crystal using an atomic frequency comb,'' {\em
  New J. Phys.}, vol.~12, no.~2, p.~023025, 2010.

\bibitem{berman2021pulsed}
P.~R. Berman and J.-L. Le~Gou\"et, ``Pulsed field transmission by atomic
  frequency combs and random spike media: The prominent role of dispersion,''
  {\em Phys. Rev. A}, vol.~103, p.~043723, Apr 2021.

\bibitem{Levy1992}
O.~Levy and D.~J. Bergman, ``Clausius-{Mossotti} approximation for a family of
  nonlinear composites,'' {\em Phys. Rev. B}, vol.~46, pp.~7189--7192, 1992.

\bibitem{Mohr2015}
P.~J. Mohr and W.~D. Phillips, ``Dimensionless units in the {SI},'' {\em
  Metrologia}, vol.~52, pp.~40--47, 2015.

\bibitem{zangwill2013modern}
A.~Zangwill, {\em Modern Electrodynamics}.
\newblock Cambridge University Press, 2013.
\newblock p.~789.

\bibitem{Beach1990}
R.~Beach, M.~D. Shinn, L.~Davis, R.~W. Solarz, and W.~F. Krupke, ``Optical
  absorption and stimulated emission of neodymium in yttrium orthosilicate,''
  {\em IEEE J. Quant. Elect.}, vol.~26, pp.~1405--1412, 1990.

\bibitem{Martin1974}
R.~M. Martin, ``Comment on the calculation of electric polarization in
  crystals,'' {\em Phys. Rev. B}, vol.~9, pp.~1998--1999, 1974.

\bibitem{Levine1991}
Z.~H. Levine and D.~C. Allan, ``Quasiparticle calculation of the dielectric
  response of silicon and germanium,'' {\em Phys. Rev. B}, vol.~43,
  pp.~4187--4207, 1991.

\bibitem{Resta1994}
R.~Resta, ``Macroscopic polarizability in crystalline dielectrics,'' {\em Rev.
  Mod. Phys.}, vol.~66, pp.~899--915, 1995.

\bibitem{Sonajalg1994}
H.~S{\~o}najalg and P.~Saari, ``Diffraction efficiency in space-and-time-domain
  holography,'' {\em J. Opt. Soc. Am. B}, vol.~11, pp.~372--379, 1994.

\bibitem{Unanyan1998}
R.~Unanyan, M.~Fleischhauer, B.~W. Shore, and K.~Bergmann, ``Robust creation
  and phase-sensitive probing of superposition states via stimulated raman
  adiabatic passage (stirap) with degenerate dark states,'' {\em Optics
  Communications}, vol.~155, pp.~144--154, 1998.

\bibitem{Grynberg2010}
G.~Grynberg, A.~Aspect, and C.~Fabre, {\em Introduction to quantum optics: from
  the semi-classical approach to quantized light}.
\newblock Cambridge University Press, 2010.

\bibitem{Plenio1998}
M.~B. Plenio and P.~L. Knight, ``The quantum-jump approach to dissipative
  dynamics in quantum optics,'' {\em Rev. Mod. Phys.}, vol.~70, pp.~101--144,
  1998.

\bibitem{Johanson2012}
J.~R. Johansson, P.~D. Nation, and F.~Nori, ``{QuTiP}: An open-source {P}ython
  framework for the dynamics of open systems,'' {\em Comp. Phys. Comm.},
  vol.~183, pp.~1760--1772, 2012.

\bibitem{attal2018rate}
Y.~Attal, P.~Berger, L.~Morvan, P.~Nouchi, D.~Dolfi, T.~Chaneli{\`e}re, and
  A.~Louchet-Chauvet, ``Rate equation reformulation including coherent
  excitation: application to periodic protocols based on spectral
  hole-burning,'' {\em JOSA B}, vol.~35, no.~6, pp.~1260--1270, 2018.

\bibitem{Longdell2002}
J.~J. Longdell, M.~J. Sellars, and N.~B. Manson, ``Hyperfine interactions in
  ground and excited states of praseodymium-doped yttrium orthosilicate,'' {\em
  Phys. Rev. B}, vol.~66, p.~035101, 2002.

\bibitem{Lovric2012}
M.~Lovri{\'c}, P.~Glasenapp, and D.~Suter, ``Spin {H}amiltonian
  characterization and refinement for {Pr}$^{3+}$:{YAlO}$_3$ and
  {Pr}$^{3+}$:{Y}$_2${SiO}$_5$,'' {\em Phys. Rev. B}, vol.~85, p.~014429, 2012.

\bibitem{Bartholomew2016}
J.~G. Bartholomew, R.~L. Ahlefeldt, and M.~J. Sellars, ``Engineering closed
  optical transitions in rare-earth ion crystals,'' {\em Phys. Rev. B},
  vol.~93, p.~014401, 2016.

\bibitem{Lapack1999}
E.~Anderson, Z.~Bai, C.~Bischof, S.~Blackford, J.~Demmel, J.~Dongarra,
  J.~Du~Croz, A.~Greenbaum, S.~Hammarling, A.~McKenney, and D.~Sorensen, {\em
  {LAPACK} Users' Guide}.
\newblock Philadelphia, PA: Society for Industrial and Applied Mathematics,
  third~ed., 1999.

\bibitem{Weiner1990a}
A.~M. Weiner and D.~E. Leaird, ``Generation of terahertz-rate trains of
  femtosecond pulses by phase-only filtering,'' {\em Opt. Lett.}, vol.~15,
  pp.~51--53, 1990.

\bibitem{Slattery2021}
K.~Slattery and Z.~H. Levine, ``Simulating photon echoes for quantum memory.''
  https://\discretionary{}{}{}demonstrations.wolfram.com/SemiclassicalSimulationOfPhotonEchoesForAtomicFrequency\discretionary{-}{}{}Comb;
  https://data.nist.gov/od/id/mds2-2454, 2021.

\end{thebibliography}

\end{document}